\documentclass[12pt]{article}
\pdfoutput=1
\usepackage{tikz}
\usepackage{dsfont}
\usepackage{amsmath}
\usepackage{amssymb}
\usepackage{braket}
\usepackage{commath}
\usepackage{float}
\usepackage{appendix}
\usepackage{fancyvrb}
\usepackage[super]{natbib}
\usepackage{times}
\usepackage{subcaption}

\title{The principle underlying antiaromaticity}

\author
{Raphael J. F. Berger$^{1\ast}$ and Alexandre Viel$^2$\\
\\
\normalsize{$^{1}$Department for Chemistry and Physics of Materials, University of Salzburg,}\\
\normalsize{Jakob-Haringerstr 2a, A-5020 Salzburg, AUSTRIA}\\
\normalsize{$^{2}$8 Sente de la haie Saint-Marc, 78480 Verneuil sur Seine, FRANCE}\\
\\
\normalsize{$^\ast$To whom correspondence should be addressed; E-mail:  raphael.berger@sbg.ac.at.}
}

\date{}

\RecustomVerbatimCommand{\VerbatimInput}{VerbatimInput}%
{fontsize=\footnotesize,
 frame=lines,  
 framesep=2em, 
 rulecolor=\color{gray},
 label=\fbox{\color{black}data.txt},
 labelposition=topline
}

\newcommand{\beginsupplement}{%
        \setcounter{table}{0}
        \renewcommand{\thetable}{S\arabic{table}}%
        \setcounter{figure}{0}
        \renewcommand{\thefigure}{S\arabic{figure}}%
}

\begin{document} 

\maketitle 

\begin{abstract}
Aromaticity is one of the most widely used chemical concepts. Current definitions
are purely phenomenological and relate symmetry, reactive 
stability and the occurrence of molecular diamagnetic response currents. The antithetical concept of 
antiaromaticity provides a connection between the contrary properties: structural 
instability or distortion out of higher symmetry, a small HOMO---LUMO gap, 
and paramagnetic response currents. We reveal the principle that is 
underlying antiaromaticity by showing an intimate and strict symmetry induced 
relation between these properties. This principle is mathematically rigorous and can be formulated like: 
{\em First order (and related) Jahn-Teller distorted molecules out of non-isometric point groups are
prone to paramagnetic current susceptibility parallel to the main axis of symmetry.} 
We show by the exemplary cases of cyclobutadiene, cyclcooctatetraene, pentalene 
and manganese trifluoride how this principle works and discuss this new perspective on 
antiaromaticity.  
\end{abstract}

\section*{}
Aromaticity is one of the most widely used chemical concepts. It aims at an abstraction of  
different experimentally observed properties in a class of chemical compounds. These properties are 
typically  a high structural or energetic stability, proneness to specific chemical reactions and 
susceptibility to induced diamagnetic currents when at the same time a theoretical description or also experimental 
evidence suggests a specific type of electronic structure. One often encounters this in the form of electron 
count rules, orbital occupation patterns or electronic resonance \cite{Miquel_2018}.
A good overview about the state of research on aromaticity including references to a selection of current reviews 
is given in the introduction of ref.\cite{10.1002/cphc.201800364}
While the connection of some of these properties seems obvious, no 
concise connection between all in full generality could ever be established. 
Here we analyze the antithetical concept of {\em anti}aromaticity which was originally 
introduced by Breslow\cite{doi:10.1021/ar50072a001} and find an intimate 
symmetry relation between its defining properties that so far seems to have escaped the general 
attention. Along the lines of the definition of 
antiaromaticity from the Gold Book of the International Union of Pure and Applied Chemistry 
(IUPAC)\cite{10.1351/goldbook.AT06987,10.1351/pac199971101919} 
these are the three properties a) ``$\dots$ prone to 
reactions causing changes in their structural type, and display tendency to alternation 
of bond lengths and fluxional behavior $\dots$'', b) ``$\dots$ a small 
energy gap between their highest occupied and lowest unoccupied molecular orbitals $\dots$'' 
and c) ``$\dots$ an external magnetic field induces a paramagnetic 
electron current.'' A full quote of the definition is given in the Supplementary Information section. 
To show how these properties are related we use the following arguments

\begin{itemize}
\item Paramagnetic molecular response is determined by virtual electronic excitations of rotational symmetry, where the axis of 
  rotation is parallel to the magnetic field.
\item Ground state---excited state (or HOMO---LUMO) symmetries that give rise to virtual electronic excitations of rotational 
  symmetry are arising from certain Jahn-Teller (JT) distortions out of specific molecular point group symmetries.
\item These certain JT distortions are first order JT distortions (or closely related ones, which we later
  call {\em primoid} second order JT distortions) and happen out of point group symmetries 
  that are non-isometric. That is from one of the point groups $C_n$, $C_{nv}$, $C_{nh}$, $D_n$, 
  $D_{nh}$ for $n>2$, and $D_{nd}$ and $S_{2n}$ for $n>1$. 
\end{itemize}
A consequence of these arguments and our main result is 
  \begin{center}
    {\em First order and primoid second order JT distorted molecules out of non-isometric point groups are 
      prone to paramagnetic current susceptibility parallel to the main axis of symmetry.}
  \end{center}

We then advocate the view that this is the symmetry principle underlying antiaromaticity. After derivation of 
these results we discuss some exemplary cases and important implications of this new perspective on 
antiaromaticity for past and future research and a possible extension to the concept of aromaticity.\\

The occurrence of paramagnetic response currents is one of the three central defining properties of 
antiaromatic molecules. We start the results section with a derivation of a symmetry based ``selection rule'' 
for virtual electronic excitations determining paramagnetic response currents.

\section*{Results}
\subsection*{Paramagnetic response currents}

When a molecule is exposed to a magnetic field $\mathbf{B}$, that is in the simplest case static, homogeneous and weak, 
which we assume throughout this work, it interacts to first order via the total electronic angular 
momentum and total spin which correspond to permanent molecular magnetic moments. In case the expectation values of 
both are identical zero, 
{\em i.e.} closed shell singlet molecules, it only interacts to second order via its molecular magnetisability which is 
for bulk mater also called magnetic susceptibility. In the quasi-classic Bohr model of electrons circling 
around the nucleus such a field dependent magnetization corresponds to a superimposed precession motion of the 
orbiting electrons in the field $\mathbf{B}$. While a magnetization in this quasi classical picture must lead to an 
increase of the energy of the electronic system and hence is classified as diamagnetic response, 
in the quantum mechanical case the magnetic response in general can be partitioned into either diamagnetic or 
paramagnetic contributions. Such a paramagnetic magnetization has no quasi-classical counterpart, it 
however would correspond to a $\mathbf{B}$ field dependent gain in angular momentum, leading to an energetic 
stabilization of the system. 
Also the partitioning into dia- and 
paramagnetic response contribution in general depends on the choice of the gauge, or in the particular case of 
a vector potential $\mathbf{A}$ of the form $\mathbf{A}_{\mathbf{d}}(\mathbf{r}) = \mathbf{(r-d)}\times\mathbf{B}$ 
on the choice of the gauge origin $\mathbf{d}$. A natural choice for the gauge origin, 
however is the origin of reference $\mathbf{r_0}$ for the definition of the angular momentum (operator) 
$\mathbf{l}=(\mathbf{r}-\mathbf{r}_0)\times\mathbf{p}$, by the momentum (operator) $\mathbf{p}$, which serves 
also as centre of symmetry in case of non-trivial molecular symmetries and conveniently can be used as 
origin of the inner coordinate system of the molecule (then $\mathbf{r}_0 = \mathbf{0}$). According to McWeeny's 
perturbation theoretic formulation\cite{10.1016/S0079-6565(99)00021-7} and orienting a unit strength 
magnetic field along the $z$ axis: $\mathbf{B}=(0,0,1)^T=\mathbf{B}_z$, the paramagnetic part of the magnetically induced 
current density field $\mathbf{J}^{\mathbf{B}_z}_p(r)$ is obtained by
\begin{equation}
 \mathbf{J}_\mathrm{p}^{\mathbf{B}_z}(\mathbf{r}) =  -\frac{2ne}{m_e} \Re{\int d\mathbf{s}_1 d\mathbf{x}_2\dots d\mathbf{x}_n {\Psi^{\mathbf{B}_z}}^{*}(\mathbf{x}_1,\dots,\mathbf{x}_n) \mathbf{p}\; {\Psi_0}(\mathbf{x}_1,\dots,\mathbf{x}_n)} \label{J_p}
\end{equation}

with the $n$ the number of electrons, the electronic charge $-e$, the electron mass $m_e$, and the ground 
state wave function $\Psi_0$ that depends on spatial coordinates of the $i$-th electron $\mathbf{r}_i$ 
and its spin coordinate $\mathbf{s}_i$ which are combined to the space-spin coordinate  $\mathbf{x}_i$ and 
the (purely imaginary) first order magnetic response contribution to the wave function 
  
\begin{equation}
\Psi^{\mathbf{B}_z}(\mathbf{x}_1,\dots,\mathbf{x}_n) = -\frac{e}{2m} \sum_{j>0} \Psi_j \frac{\Braket{\Psi_j|\mathbf{l}_z|\Psi_0}}{\varepsilon_j - \varepsilon_0},\label{L}
\end{equation}
using $\mathbf{l}_z$ for the $z$-component of the (total) angular momentum operator
(alternatively an effective one-particle picture formulation can be easily obtained, 
resulting in an completely analogous expression for occupied-unoccupied orbital excitations). 
Thus the possibility of occurrence of a paramagnetic response contribution  
$\mathbf{J}_\mathrm{p}^{\mathbf{B}_z}(\mathbf{r})$ is determined by the condition

\begin{equation}
 \frac{\abs{\Braket{\Psi_j|\mathbf{l}_z|\Psi_0}}}{\varepsilon_j - \varepsilon_0} > 0,\label{vt}
\end{equation} 

which in turn can be interpreted as a selection rule for $\mathbf{J}_\mathrm{p}^{\mathbf{B}_z}(\mathbf{r})$ requiring 
the possibility of virtual excitations between $\Psi_0$ and $\Psi_j$ of rotational symmetry with respect to an axis parallel to 
$z$, which is the direction of the magnetic field $\mathbf{B}_z$. The connection between paramagnetism and rotation 
to the best of our knowledge was pointed out for the first time by Steiner \& Fowler\cite{10.1039/B104847N}. 
According to the Wigner-Eckart theorem we can  
formulate the selection rule, that the decomposition of the direct (tensor) product of the 
irreducible representations (IRs, $\Gamma$) 
of the ground and excited states must contain the IR of the $z$-component of the angular momentum operator:

\begin{equation}
 \Gamma_{\mathbf{l}_z} \subseteq \Gamma_{\Psi_j^*}\otimes\Gamma_{\Psi_0}. \label{para_rule}
\end{equation} 

The terms and hence total paramagnetic currents in general will be the stronger the 
larger the integrals $\abs{\Braket{\Psi_j|\mathbf{l}_z|\Psi_0}}$ and the smaller the energy gaps $\varepsilon_j - \varepsilon_0$ 
between the engaged states are. 

In this specific choice of gauge, that is a common gauge origin, one can introduce the terms 
dia- and paratropic\cite{C1CP22457C}, which are referring to ring current contributions to the total 
current density vector field that are 
flowing either in classical direction or anti-classical direction around the gauge origin and with respect to the 
external magnetic field, thus giving rise to a dia- or paramagnetic induced magnetisation inside this ring 
current domain, respectively.

Representations of angular momentum components play now a central role in our considerations. In the following we will see 
that there is an intimate connection between tensor squares of two-dimensional IRs and representations of 
angular momentum components in certain point groups.
  
\subsection*{Angular momentum in squares of degenerate IR$s$} 

For here on we adopt the Mulliken conventions of $A/a$, and $B/b$ for one-dimensional, $E/e$ for two-dimensional, 
$T/t$ for three-dimensional irreducible representations over $\mathbb{R}$, $\rho$ for a general representation, $\Gamma$ 
for an irreducible one and $\Gamma_0$ for the trivial (= totally symmetric) representation. Also in this whole work we 
consider only representations over $\mathbb{R}$.\\

Classifing the point groups by the dimensions of their IRs yields a partition 
into three subfamilies that we call $\mathcal{K}_{aaa}$, $\mathcal{K}_{ae}$ and $\mathcal{K}_{t}$.
In informal chemical terminology point groups from $\mathcal{K}_{aaa}$ are those which 
only have one-dimensional IRs (over $\Bbb R$, as used throughout), point groups from 
$\mathcal{K}_{ae}$ are those which have two- but no higher dimensional IRs and point groups from 
$\mathcal{K}_{t}$ have three- (or higher) dimensional IRs.\\

\begin{table}[H]
\caption{Angular momentum and $E\otimes E$ representation tables for all point groups \label{TABLE}} 
\label{alt2table}

\begin{subtable}{0.45\textwidth}
\caption{$\mathcal{K}_{aaa}$ point groups} 
\label{Kaaa}
\begin{tabular}{|c|cccccccc|}
\hline
p.grp & $C_1$ & $C_s$ & $C_2$ & $C_i$ & $C_{2v}$ & $C_{2h}$ & $D_2$ & $D_{2h}$ \\
\hline
$\Gamma_{\mathbf{l}_z}$ & $A$ & $A'$ & $A$ & $A$ & $A_2$ & $A_g$ & $B_1$ & $B_{1g}$ \\
\hline
\end{tabular}
\end{subtable}
\vspace{0.75cm}

\begin{subtable}{0.45\textwidth}
\caption{$\mathcal {K}_{ae}$ cyclic-based point groups} 
\label{Kaecyc}
\begin{tabular}{|c|c|c|c|c|c|c|c|c|}
\hline
p.grp, $n \ge 1$ & $C_{2n+1}$ & $C_{2n+2}$ & $C_{(2n+1)v}$ & $C_{(2n+2)v}$ & $C_{(2n+1)h}$ & $C_{(2n+2)h}$ & $S_{4n}^{(*)}$ & $S_{4n+2}$ \\
\hline
$\Gamma_e$, $i=1 \ldots n$ & $E_i$ & $E_i$ & $E_i$ & $E_i$ & $E_i', E_i''$ & $E_{ig},E_{iu}$ & $E_i$ & $E_{ig}, E_{iu}$ \\ 
\hline
$\Gamma_{\mathbf{l}_z}$ & \multicolumn{2}{c|} {$A$ } & \multicolumn{2}{c|} {$A_2$ } & $A'$ & $A_g$ & $A$ & $A_g$ \\
\hline
$[\Gamma\otimes\Gamma]$ & \multicolumn{2}{c|} {$A$ } & \multicolumn{2}{c|} {$A_2$ } & $A'$ & $A_g$ & $A$ & $A_g$ \\
\hline
\end{tabular}
$^{(*)}$ : $i$ ranges from $1$ to $2n-1$
\end{subtable}
\\
\vspace{0.75cm}

\begin{subtable}{0.45\textwidth}
\caption{$\mathcal {K}_{ae}$ dihedral-based point groups}
\label{Kaedih}
\begin{tabular}{|c|c|c|c|c|c|c|}
\hline
p.grp, $n \ge 1$ & $D_{2n+1}$ & $D_{2n+2}$ & $D_{(2n)d}^{(*)}$ & $D_{(2n+1)d}$ &  $D_{(2n+1)h}$ & $D_{(2n+2)h}$ \\
\hline
$\Gamma_e$, $i=1 \ldots n$ & $E_i$ & $E_i$ & $E_i$ & $E_{ig}, E_{iu}$ & $E_i',E_i''$ & $E_{ig}, E_{iu}$ \\ 
\hline
$\Gamma_{\mathbf{l}_z}$ 
& \multicolumn{2}{c|} {$A_2$} 
& $A_2$ & $A_{2g}$
& $A_2'$ & $A_{2g}$
 \\
\hline
$[\Gamma\otimes\Gamma]$ 
& \multicolumn{2}{c|} {$A_2$} 
& $A_2$ & $A_{2g}$
& $A_2'$ & $A_{2g}$
 \\
\hline
\end{tabular}
$^{(*)}$ : $i$ ranges from $1$ to $2n-1$
\end{subtable}
\\
\vspace{0.75cm}

\begin{subtable}{0.45\textwidth}
\caption{$\mathcal{K}_t$ type cubic point groups}
\label{Ktcub}
\begin{tabular}{|c|cc|ccc|cccc|cccccc|}
 \hline
p.grp 
& \multicolumn{2}{c|}{ $T$ }
& \multicolumn{3}{c|}{ $T_d = O$ }
& \multicolumn{4}{c|}{ $T_h$ }
& \multicolumn{6}{c|}{ $O_h$ }
\\
\hline
$\Gamma_l$
& \multicolumn{2}{c|}{ $T$ }
& \multicolumn{3}{c|}{ $T_1$ }
& \multicolumn{4}{c|}{ $T_g$ }
& \multicolumn{6}{c|}{ $T_{1g}$ }
\\
\hline
 $\Gamma$ & $E$ & $T$
 & $E$ & $T_1$ & $T_2$
& $E_g$ & $T_g$ & $E_u$ & $T_u$
& $E_g$ & $T_{1g}$ & $T_{2g}$ & $E_u$ & $T_{1u}$ & $T_{2u}$
\\
 \hline
 $[\Gamma\otimes\Gamma]$ & $A$ & $T$ 
 & $A_2$ & $T_1$ & $T_1$ 
& $A_g$ & $T_g$ & $A_g$ & $T_g$ 
& $A_{2g}$ & $T_{1g}$ & $T_{1g}$ & $A_{2g}$ & $T_{1g}$ & $T_{1g}$
 \\
 \hline
\end{tabular}
\end{subtable}
\vspace{0.75cm}

\begin{subtable}{0.45\textwidth}
\caption{$\mathcal{K}_t$ type icosahedral groups} 
\label{Ktico}
\begin{tabular}{|c|c|c|c|c|}
 \hline
p.grp 
& \multicolumn{4}{c|}{ $I$ }
\\
\hline
$\Gamma_\mathbf{l}$
& \multicolumn{4}{c|}{ $T_1$ }
\\
\hline
 $\Gamma$ & $T_1$ & $T_2$ & $G$ & $H$
\\
 \hline
 $[\Gamma\otimes\Gamma]$ & $T_1$ & $T_2$ & $T_1\oplus T_2$ & $T_1\oplus T_2\oplus G$ 
 \\
 \hline
\end{tabular}
{}\\

\begin{tabular}{|c|c|c|c|c|c|c|c|c|}
 \hline
p.grp 
& \multicolumn{8}{c|}{ $I_h$ }
\\
\hline
$\Gamma_\mathbf{l}$
& \multicolumn{8}{c|}{ $T_{1g}$ }
\\
\hline
 $\Gamma$ & $T_{1g}$ & $T_{2g}$ & $G_g$ & $H_g$ & $T_{1u}$ & $T_{2u}$ & $G_u$ & $H_u$
\\
 \hline
 $[\Gamma\otimes\Gamma]$ 
& $T_{1g}$ & $T_{2g}$ & $T_{1g} \oplus T_{2g}$ & $T_{1g} \oplus T_{2g}\oplus G_g$ 
& $T_{1g}$ & $T_{2g}$ & $T_{1g} \oplus T_{2g}$ & $T_{1g}\oplus T_{2g}\oplus G_g$ 
 \\
 \hline
\end{tabular}
\end{subtable}
\end{table}

We now put our focus on the symmetry properties of the quantum mechanical total angular momentum operator $\mathbf{l}$ and 
its $z$-component $\mathbf{l}_z$  in the three point group families. A 
comparison with Table \ref{alt2table} shows that for all two-dimensional IRs $E$ for all groups 
$G \in \mathcal{K}_{ae}$ the decompostion of the tensor square of $E$ into irreducible representations 
always contains $\Gamma_{\mathbf{l}_z}$, namely in its anti-symmetric part. In particular
\begin{equation}
  E \otimes E = \Gamma_{\mathbf{l}_z} \oplus \Gamma_0 \oplus \rho_q;\;\; \dim(\Gamma_0) = \dim(\Gamma_{\mathbf{l}_z})=1,  \dim(\rho_q)=2,
\label{E2_in_K}
\end{equation}
with
\begin{equation}
  [E \otimes E] = \Gamma_{\mathbf{l}_z}
\label{Alt2_E_in_K}
\end{equation}
holds, with $[M]$ denoting the anti-symmetric part of a tensor $M$ (below and in the SI section $[M]$ is 
identified with the so called ``alternating square'' functor $Alt^2(M)$).\\

On the contrary, for groups from $\mathcal{K}_{t}$, $[E\otimes E]$ does not represent (any components) of $\mathbf{l}$ 
(see Table \ref{alt2table}). 
Rigorous definitions and reasoning, the relation between $\mathbf{l}$ and the representation space $O(3)$ and 
further detailed outlines are given in the SI section.\\

So we see that in point groups from $\mathcal{K}_{ae}$ there is an 
intimate connection between two-dimensional IRs and 
$\Gamma_{\mathbf{l}_z}$. In the next section we consider 
branchings of two-dimensional electronic levels at a group theoretical level.

\subsection*{Distortions}

Distorting a molecule from a higher to a lower symmetric structure leads to a
restriction of the representations of molecular eigenstates to a subgroup $H$ of $G$, which
can change the shape of its irreducible decomposition. This is usually called 
'descent in symmetry' in chemistry, the mathematical discipline concerned with this is 
'branching representation theory'.\\

Consider a point group $\mathcal H \subset \mathcal G$, and a two dimensional representation 
$E$ of $\mathcal G$. If the restriction of $E$ to $\mathcal H$ (= $E|_\mathcal{H}$) branches 
into a direct sum of two one-dimensional representations of $\mathcal H$,
\begin{equation} 
E|_{\mathcal{H}} = A_\alpha \oplus A_\beta
\end{equation} 
then 
\begin{equation} 
[E\otimes E]|_{\mathcal H} = [E|_{\mathcal H} \otimes E|_{\mathcal H}] = A_\alpha \otimes A_\beta
\end{equation} 
as is lined out in the SI section in subsection \ref{S1}. If $E$ was the representation of the point 
group $\mathcal G$ on the $(x,y)$-plane, then
\begin{equation} 
\Gamma_{\hat l_z}|_{\mathcal H} =  [E\otimes E]|_{\mathcal H}  =  A_\alpha \otimes A_\beta.
\label{angmom}
\end{equation} 
The angular momentum representation of $\mathcal H$ on the rotations around the $z$-axis is the tensor product of the two representations $A_\alpha, A_\beta$.\\

This implies that if an $E$-type IR in a non-isometric point group is restricted to a subgroup such that it splits
into two non-degenerate hence one-dimensional IRs, then the direct product of these two one-dimensional 
representations is exactly the representation of the z-component of the angular momentum operator (in the subgroup). This is an 
essential result.\\

In particular the Jahn-Teller (JT) distortions to be discussed in the next section can trivially be 
described as a descent in symmetry, such that all considerations from above are valid for JT distortions.

\subsection*{'Primoid' second order JT systems}

We adopt in the following a simplistic view on the Jahn-Teller-Effect based on the original 
idea from Landau and Teller where distortions were considered in the form of small perturbations and based on 
a specific situation of electronic degeneracy. As a starting point for a deeper 
understanding and a contemporary view on the JT-effect we refer the reader to I. B. Bersuker's introductory 
text ``Recent Developments in the Jahn-Teller Effect theory''\cite{10.1007/978-3-642-03432-9}.
We are focusing only on the basic integrals occurring in the perturbation expansion\cite{10.1002/9781118558409} 
of the molecular energy with respect to a deviation (or distortion) $Q$ 
from the minimum geometry on the adiabatic potential energy surface. In 
general $Q$ may be regarded as some symmetry adapted local nuclear 
distortion coordinate, like a vibrational normal modes. 
In that sense the occurrence of a first order JT distortion is 
determined by the size of the first order coefficients in $Q$:

\begin{equation}
\braket{\Psi_0|\frac{d \hat{H}}{d Q}|\Psi_0} \label{c1}
\end{equation}
with the ground state wave function $\Psi_0$, while second order JT distortions are 
controlled by the negative components of the second order 
coefficients\cite{10.1002/9781118558409}
\begin{equation}
\sum_{i>0}\frac{\abs{\braket{\Psi_0|\frac{\partial\hat{H}}{\partial Q}|\Psi_i}}^2}{\varepsilon_0^0 - \varepsilon_i^0} \label{c2}
\end{equation}
with the virtual state wave functions $\Psi_i$, and the corresponding unperturbed state energies $\varepsilon^0_j$ (for $j=0,i$).\\

Let us consider now a one electron (for simplicity) molecule in a twofold degenerate, $E$ type, one-electron 
state $\psi_0$, then the JTT implies that for a non-linear point group symmetry $G$ 
a non-totally symmetric distortion mode $Q$ exists such that according to expression \ref{c1} we have 
$\Gamma_Q \subseteq\Gamma_{\psi_0}^* \otimes \Gamma_{\psi_0}$. \\

When we now interpret this one-electron $E$ state as a spatial orbital from which we can built  
two-electron configuration state functions from symmetry adapted linear combinations of 
Slater determinants of these (spin)orbitals, we arrive at various possible combinations of ground and virtual 
state ($\Psi_0$ and $\Psi_i$) symmetries:  
$\Gamma_{\Psi_0}, \Gamma_{\Psi_i} \subset \Gamma_{\psi_0} \otimes \Gamma_{\psi_0} = \Gamma_0 \oplus Alt^2(E) \oplus \rho_q$ (with eq. \ref{E2_in_K}). 
From those one can choose $\Gamma_{\Psi_0}=\Gamma_0$ and $\Gamma_{\Psi_i}=\Gamma_Q$, 
yielding, as the Wigner-Eckart theorem applied to expression \ref{c2} suggests
$\Gamma_Q \subseteq\Gamma_{\psi_0}^* \otimes \Gamma_{\psi_i} = \Gamma_0  \otimes \Gamma_Q = \Gamma_Q$
, such that a 
second order JT distortion by the mode $Q$ is possible. Such a choice we call 
``{\bf primoid} second order JT system''. In the practice of quantum chemical calculations, 
primoid second order JT systems reveal their nature in the orbital space from which their state functions 
can be constructed. This is demonstrated in the degeneracy and branching of orbital energies like in C$_4$H$_4$ 
(see below and Fig. \ref{C4H4_levels}).\\

Also, by this two-electron wave function construction, from an one-electron wave function, 
both $\Psi_0$ and the virtual state $\Psi_i$ can only be in the decomposition of $E \otimes E$, 
which together with the restriction to cases were JT causes only small structural 
perturbations can warrant that the energy denominator in expression \ref{c2} is not too large. \\

For a counter example of a {\em non-primoid} second order JT case \label{non-primoid}, consider Si$_4$F$_4$ in the 
abelian D$_{2h}$ symmetry\cite{10.1038/srep23315} which distorts via a $B_{1g}$ mode 
to C$_{2h}$, or similarly Si$_2$Ge$_2$F$_4$ which distorts from D$_{2h}$ via C$_{2h}$ to C$_{s}$. 
In general, all cases of second order Jahn-Teller distortion which are of a reference 
(or highest possible) point group symmetry that possesses no degenerate IR$s$ over ${\Bbb R}$, 
are non-primoid second order Jahn-Teller cases.\\

\subsection*{JT and paramagnetism in point groups from $\mathcal{K}_{ae}$}

All point groups $G$ in the family $\mathcal{K}_{ae}$ (see Table \ref{TABLE}) are represented by at least one two-dimensional IR over ${\Bbb R}$, 
say $E$ and in these point groups $Alt^2(E)$ represents for all $E$ the $z$-component of the quantum mechanical angular 
momentum operator $\mathbf{l}_z$. Now molecular systems in 
electronic states represented by $E$ are due to their degeneracy prone to a first order JT distortion. 
If such a distortion to a subgroup $H$ of $G$ occurs then $E$ branches into two one-dimensional representations, 
say $A_\alpha$ and $A_\beta$. The corresponding lower lying electronic state say $A_\alpha$ eventually will 
become the ground state (or a SOMO or HOMO orbital in the effective one-particle picture) while the higher lying state
$A_\beta$ will become some virtual state (or a UMO or possibly the LUMO depending on the size of the distortion). Then and especially 
if the distortion is not too large, such that the energy difference between occupied and virtual states will not become too large,
the system becomes by virtue of the symmetry selection rule for paratropic currents
(eqns. \ref{J_p} and \ref{vt}) strongly susceptible to induced paratropic currents (since according to eq. \ref{angmom}
$A_\alpha\otimes A_\beta=\Gamma_{\mathbf{l}_z}$). In this way we see that all first order JT 
cases arising from  point groups $G$ in the family $\mathcal{K}_{ae}$, that is all non-isometric first oder JT cases, are inherently prone 
to paramagnetism.\\

Since this result is based on group theoretical considerations it can be seen as a completely 
general symmetry property that combines structure and magnetic response. Similar like {\em e.g.} 
the dipole transition rule which gives a general symmetry property of the molecular response 
with respect to any dipole fields. Like such spectroscopic selection rules, the predictivity 
is of course limited by the fact that we deal merely with symmetry rules which are completely 
independent of the peculiar quantitative electronic properties of a molecule under investigation. 
But also similar to these selection rules the principle might serve as a basis for the understanding 
of the phenomenon of antiaromaticity in a very general sense.\\

It is very important to note that these considerations are not restricted to the first order JT effect since this is bound to open shell systems,
and the most important examples for antiaromaticity like 1,3-cyclobutadiene are closed shell cases. 
From the definition of the {\em primoid} second order JT case it follows immediately that there is the very same connection between
symmetry and paratropic response currents. One example is 1,3-cyclobutadiene, that goes back to states derived 
from a doubly occupied doubly degenerate orbital.\\

We now can formulate the main result of this work, as is announced in the introduction:\\

    {\em First order and primoid second order JT distorted molecules out of non-isometric point groups are 
      prone to paramagnetic current susceptibility parallel to the main axis of symmetry.}\\

Assuming small distortions, it is clear that these cases match the complete IUPAC definition of antiaromaticity 
(see Introduction and SI) such that a strict connection between the defining properties is given. Since to the best of our 
knowledge there exists no undoubted example for an antiaromatic molecule that does not belong to this class, we suggest that
this is the symmetry principle underlying antiaromaticity.\\

In the following we will discuss four representative examples for molecules, 
each of them reflecting a slightly different situation in hindsight to its
antiaromatic character. 

\subsection*{Examples}

\subsubsection*{C$_4$H$_4$}

1,3-Cyclobutadiene (C$_4$H$_4$) is such a prominent example for antiaromaticity 
that it is mentioned in the IUPAC definition (see ref. 
\cite{10.1351/goldbook.AT06987,10.1351/pac199971101919} and the SI for a full 
quote). Details on its symmetry, electronic states and the nature of the Jahn-Teller 
distortion have been worked out by Nakamura \& co-workers \cite{10.1016/0301-0104(89)80129-6} 
in the framework of a MCSCF study. It is a case where for the full $D_{4h}$ 
symmetry we have double occupation of the doubly degenerate $e_g$ orbitals 
($e_g^2$), which results in purely non-degenerate electronic states, the first four of
which are in energetically ascending order $^1B_{1g}, ^3A_{2g}, ^1A_{1g}, ^1B_{2g}$. Hence,
this is a prototypical example for a primoid second order Jahn-Teller case. 

Formally removing one electron yields the [C$_4$H$_4$]$^+$ cation and a $^2E_g$ ground state. 
To check the possibility of a non-zero integral term \ref{c1}, thus 
the possibility of existence of a distortional mode $Q$ leading to 
an energetic stabilization, which is a condition for a first order 
JT effect, we need to decompose $E_g\otimes E_g$ into a direct sum of IRs. This yields
\begin{equation}
E_g\otimes E_g = A_{1g}\oplus[A_{2g}]\oplus B_{1g}\oplus B_{2g}
\label{eu2}
\end{equation}
Permissible distortional mode symmetries (as must be checked separately) here are either 
$B_{1g}$ or $B_{2g}$, these are related by a $45^\circ$ rotation around $z$, and give rise either 
to a rhombic or a rectangular $D_{2h}$ symmetric structure, respectively. Quantum chemical 
calculations show that in this case the rhombic structure is preferred\cite{10.1016/0301-0104(89)80129-6}
and leads to a splitting of the degenerate $e_g$ orbitals into two non-degenerate 
$b_{2u}$ and $b_{3u}$ orbitals.\\ 

Since $D_{4h}$ is a non-isometric point group (case $\mathcal{K}_{ae}$) and $E_g$ is 
a degenerate level we know according to eq. \ref{E2_in_K} that $E_g\otimes E_g$ 
contains the IR for the $z$-component of the angular momentum $\Gamma_{\mathbf{l}_z}$, 
which in $D_{4h}$ is $A_{2g}$. Moreover from eq. \ref{lz_in_K} follows that this is 
$\Gamma_{\mathbf{l}_z}=Alt^2(E_g)=[E_g\otimes E_g]=A_{2g}$. According to eq. \ref{angmom}
in the new group $D_{2h}$ the IR for $\mathbf{l}_z$ is contained in the direct product 
of the two branches of $E_u$. That this is indeed the case 
shows a quick check: in $D_{2h}$ we have $B_{2u}\otimes B_{3u} = B_{1g}$, which is
identical to $\Gamma_{\mathbf{l}_z}$ in this group.\\

As we have identified C$_4$H$_4$ by its orbital structure as a (potential) primoid second order JT case 
by recursion to the hypothetical [C$_4$H$_4$]$^+$ case, we could easily identify the possible symmetries 
of the $Q$ mode that could be operational in a second order JT effect. Note that since this a (primoid) 
second order JT distortion no electronic state degeneracy is involved but only a coupling between the 
$^1B_{1g}$ ground state and an $^1A_{1g}$ excited state takes place which is moderated by the $B_{1g}$ type 
distortion. The degeneracy and branching only is reflected in the underlying orbital structure which leads 
to the classification as a {\em primoid} second order JT system.
And in addition by the arguments from above on the angular momentum representation 
we know that in the ``real'' $n$-electron state case with a $(e_g)^2$ occupation there will be 
occupied-to-virtual transitions of $\Gamma_{\mathbf{l}_z}$ a symmetry species available in the accordingly 
distorted molecule.\\
 
We can verify this as well: Eq. \ref{eu2} shows all permissible two electron states from the $(e_g)^2$ occupation.
As a matter of fact the ground state turns out to be of a $^1B_{1g}$ symmetry species and out of the virtual states
given by the remaining three terms of eq. \ref{eu2} again only a virtual transition to the $^1A_{1g}$ state 
is of a symmetry species that is in accordance with a permissible distortional coordinate $Q$ (see [C$_4$H$_4$]$^+$ case), 
that is
\begin{equation}
 B_{1g} \otimes A_{1g} = B_{1g}.
\end{equation}
So both models predict of course the same possible symmetries of distortion $Q$ out of $D_{4h}$ to $D_{2h}$. To find the by virtue 
of the ``primoid'' argument predicted virtual transition of $\Gamma_{\mathbf{l}_z}$ symmetry species 
we have to restrict all terms on the right 
side of eq. \ref{eu2} to the subgroup $D_{2h}$. This gives 
\begin{equation}
(A_{1g}\oplus A_{2g}\oplus B_{1g}\oplus B_{2g})_{|D_{2h}} = A_g \oplus B_{1g} \oplus A_g \oplus B_{1g} 
\end{equation}
Since thus the ground state ``branches'' from $B_{1g}$ to $A_g$, for example a virtual transition $A_g \to B_{1g}$ is available which 
corresponds obviously to a $B_{1g}$ symmetry species and, that as we have used already above, is 
identical to $\Gamma_{\mathbf{l}_z}$ in $D_{2h}$.

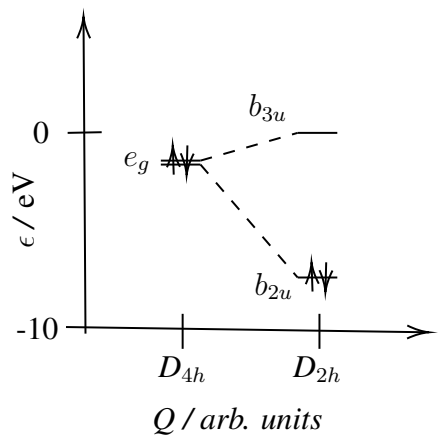
\begin{figure}
\centering
\tikzset{every picture/.style={line width=0.75pt}} 
\begin{tikzpicture}[x=0.75pt,y=0.75pt,yscale=-1,xscale=1]
\draw    (41.17,158.67) -- (222.17,161.3) ;
\draw [shift={(224.17,161.33)}, rotate = 180.83] [color={rgb, 255:red, 0; green, 0; blue, 0 }  ][line width=0.75]    (10.93,-3.29) .. controls (6.95,-1.4) and (3.31,-0.3) .. (0,0) .. controls (3.31,0.3) and (6.95,1.4) .. (10.93,3.29)   ;
\draw    (51,162) -- (50.01,4) ;
\draw [shift={(50,2)}, rotate = 449.64] [color={rgb, 255:red, 0; green, 0; blue, 0 }  ][line width=0.75]    (10.93,-3.29) .. controls (6.95,-1.4) and (3.31,-0.3) .. (0,0) .. controls (3.31,0.3) and (6.95,1.4) .. (10.93,3.29)   ;
\draw    (100,152) -- (100,169.67) ;
\draw    (169,152) -- (169,169.67) ;
\draw    (109,76.67) -- (89,76.67) ;
\draw    (178,133.67) -- (158,133.67) ;
\draw    (95.5,81.75) -- (95.07,69) ;
\draw [shift={(95,67)}, rotate = 448.06] [color={rgb, 255:red, 0; green, 0; blue, 0 }  ][line width=0.75]    (10.93,-3.29) .. controls (6.95,-1.4) and (3.31,-0.3) .. (0,0) .. controls (3.31,0.3) and (6.95,1.4) .. (10.93,3.29)   ;
\draw    (165.5,140.75) -- (165.07,128) ;
\draw [shift={(165,126)}, rotate = 448.06] [color={rgb, 255:red, 0; green, 0; blue, 0 }  ][line width=0.75]    (10.93,-3.29) .. controls (6.95,-1.4) and (3.31,-0.3) .. (0,0) .. controls (3.31,0.3) and (6.95,1.4) .. (10.93,3.29)   ;
\draw  [dash pattern={on 4.5pt off 4.5pt}]  (109,76.67) -- (158,133.67) ;
\draw    (102.4,67) -- (102.05,80) ;
\draw [shift={(102,82)}, rotate = 271.53] [color={rgb, 255:red, 0; green, 0; blue, 0 }  ][line width=0.75]    (10.93,-3.29) .. controls (6.95,-1.4) and (3.31,-0.3) .. (0,0) .. controls (3.31,0.3) and (6.95,1.4) .. (10.93,3.29)   ;
\draw    (172.4,126) -- (172.05,139) ;
\draw [shift={(172,141)}, rotate = 271.53] [color={rgb, 255:red, 0; green, 0; blue, 0 }  ][line width=0.75]    (10.93,-3.29) .. controls (6.95,-1.4) and (3.31,-0.3) .. (0,0) .. controls (3.31,0.3) and (6.95,1.4) .. (10.93,3.29)   ;
\draw    (109,74.67) -- (89,74.67) ;
\draw    (59.17,60.33) -- (42,60.67) ;
\draw    (178,60.67) -- (158,60.67) ;
\draw  [dash pattern={on 4.5pt off 4.5pt}]  (109,74.67) -- (158,60.67) ;
\draw (100,180) node  [align=left] {\textit{D}$\displaystyle _{4h}$};
\draw (168,180) node  [align=left] {\textit{D}$\displaystyle _{2h}$};
\draw (127.67,206.33) node  [align=left] {\textit{Q / arb. units}};
\draw (77,76) node   {$e_{g}$};
\draw (28.67,60.33) node  [align=left] {0};
\draw (26.67,159.33) node  [align=left] {\mbox{-}10};
\draw (146,138) node   {$b_{2u}$};
\draw (143,48) node   {$b_{3u}$};
\draw (19.67,99.33) node [rotate=-270] [align=left] {$\displaystyle \epsilon $\textit{ / }eV};
\end{tikzpicture}
\caption{Frontier orbital energies $\varepsilon$ for the ground state saddle point ($D_{4h}$) or  
minimum structure ($D_{2h}$) calculated at the generalized valence bond - perfect pairing (GVB(PP)/TZVPP) level of theory for 
cyclooctatetraene (C$_8$H$_8$). Orbital energy splitting upon distortion $Q$ from $D_{4h}$ to the $D_{2h}$ minimum is indicated 
by the dashed lines. The distortion from $D_{4h} \to D_{2h}$ is a {\em primoid} second order Jahn-Teller (JT) distortion via a $B_{1g}$ mode. 
The corresponding electronic levels are shown in Fig. \ref{C4H4_orb_branch} in the SI. Note that the electronic levels are 
{\em not degenerate} (second order JT system), but the underlying orbital structure that is depicted here shows degeneracy and branching 
({\em primoid} second order JT system). The direct products $E_{g}\otimes E_{g}$ and $B_{2u}\otimes B_{3u}$, the 
latter corresponding to the paramagnetic virtual excitation, 
contain or are the IR for the $z$-component of the angular momentum operator. The molecule in the $D_{4h}$ saddle point configuration 
shows a much stronger paramagnetic response than the $D_{2h}$ minimum structure (compare Fig. \ref{C4H4_currents}) \label{C4H4_levels} but 
both are classified as antiaromatic.}
\end{figure}

So we have shown how the argument of prediction of virtual transitions of $\Gamma_{\mathbf{l}_z}$ symmetry, that is 
paramagnetic response, for the primoid second order JT case C$_4$H$_4$ (out of non-isometric point groups) works. 
Calculations show (see SI for details) that we can expect a double bond localization, or at 
least a splitting of C-C distances from 
1.426 \AA ($D_{4h}$) to 1.552 and 1.324 \AA.
The weight of the second configuration is about 4\% at the $D_{2h}$ minimum structure, thus it is safe to use 
a single reference method for calculation of the magnetically induced ring currents (see SI for details).
The integral of the total global current susceptibility amounts to -22.4 nA/T paratropic total current at 
HF/VTZ level of theory.\\

For the aromatic benzene we find diamagnetic current contributions of about +20 nA/T. One might say C$_4$H$_4$ is as 
paramagnetic as benzene is diamagnetic (but one should note that there are paramagnetic contributions to the benzene 
currents as well and these are of a size of about -9 nA/T)\cite{10.1039/B101179K}.\\

C$_4$H$_4$ is the classical example for antiaromaticity, in agreement with that we find that its ground state geometry is derived
from a primoid second order Jahn-Teller distortion out of a point group of the family $\mathcal{K}_{ae}$, 
it has a small HOMO-LUMO gap and thus as a consequence of the latter two arguments 
shows a strong paramagnetic response with respect to a magnetic field $\mathbf{B}$ parallel to the original $C_4$ axis.

\begin{figure}[H]
\centering
\begin{subfigure}{0.9\textwidth}
\includegraphics[width=0.8\textwidth]{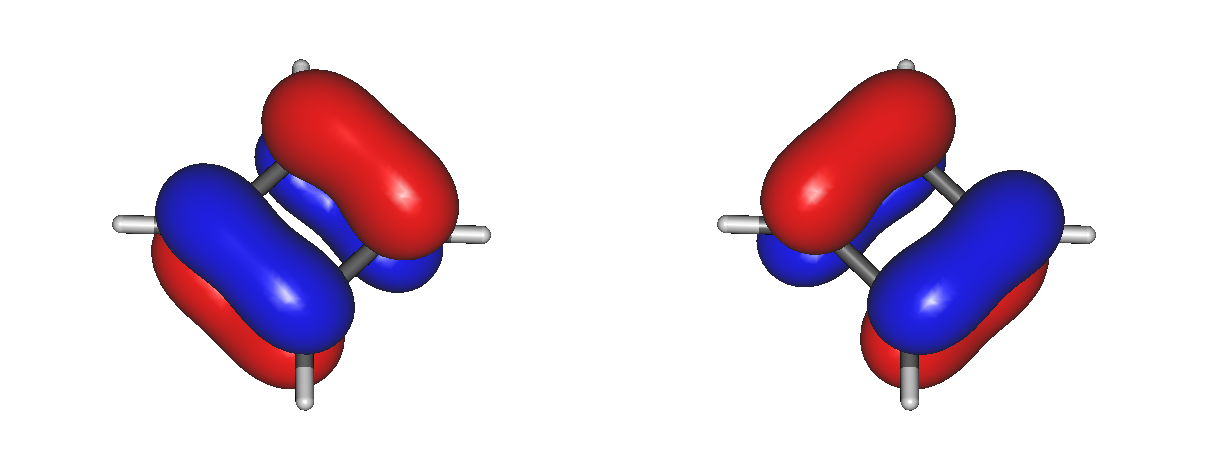}
\caption{Highest occupied molecular orbital (HOMO) in the $D_{4h}$ saddle point structure. The two-fold degenerate orbital is represented by the $e_{g}$ IR.}
\end{subfigure}
\begin{subfigure}{0.9\textwidth}
\includegraphics[width=0.8\textwidth]{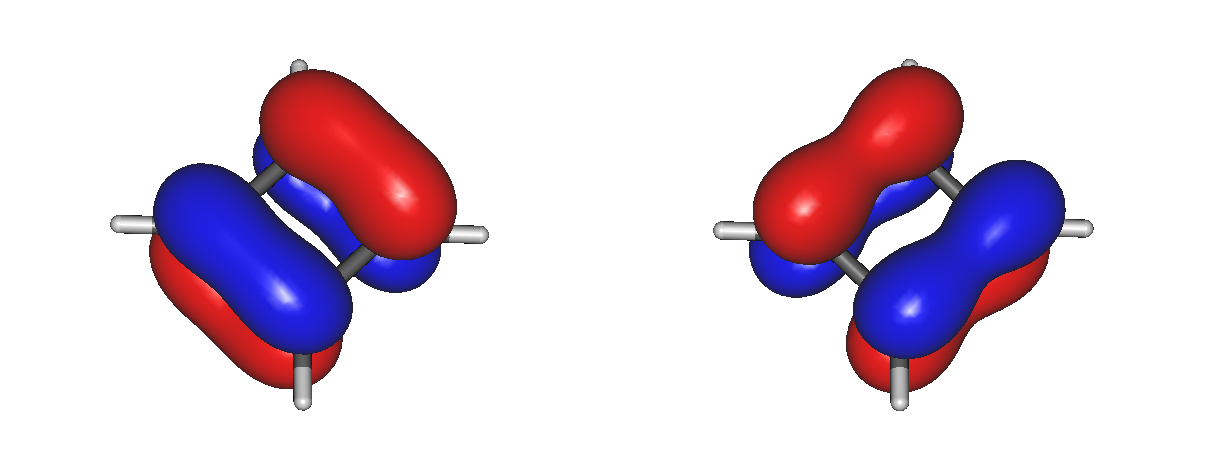}
\caption{HOMO (left) and lowest unoccupied molecular orbital (LUMO, right) in the $D_{2h}$ minimum structure. The HOMO is represented by $b_{2u}$, the LUMO by the $b_{3u}$ IR.}
\end{subfigure}
\caption{Frontier orbitals of the electronic ground state in C$_4$H$_4$ at different symmetries, calculated at the GVB(PP)/TZVPP level of theory, displayed as 
isosurfaces at densities of 0.066.\label{C4H4-orbitals}}
\end{figure}

\begin{figure}[H]
\centering
\includegraphics[width=17cm]{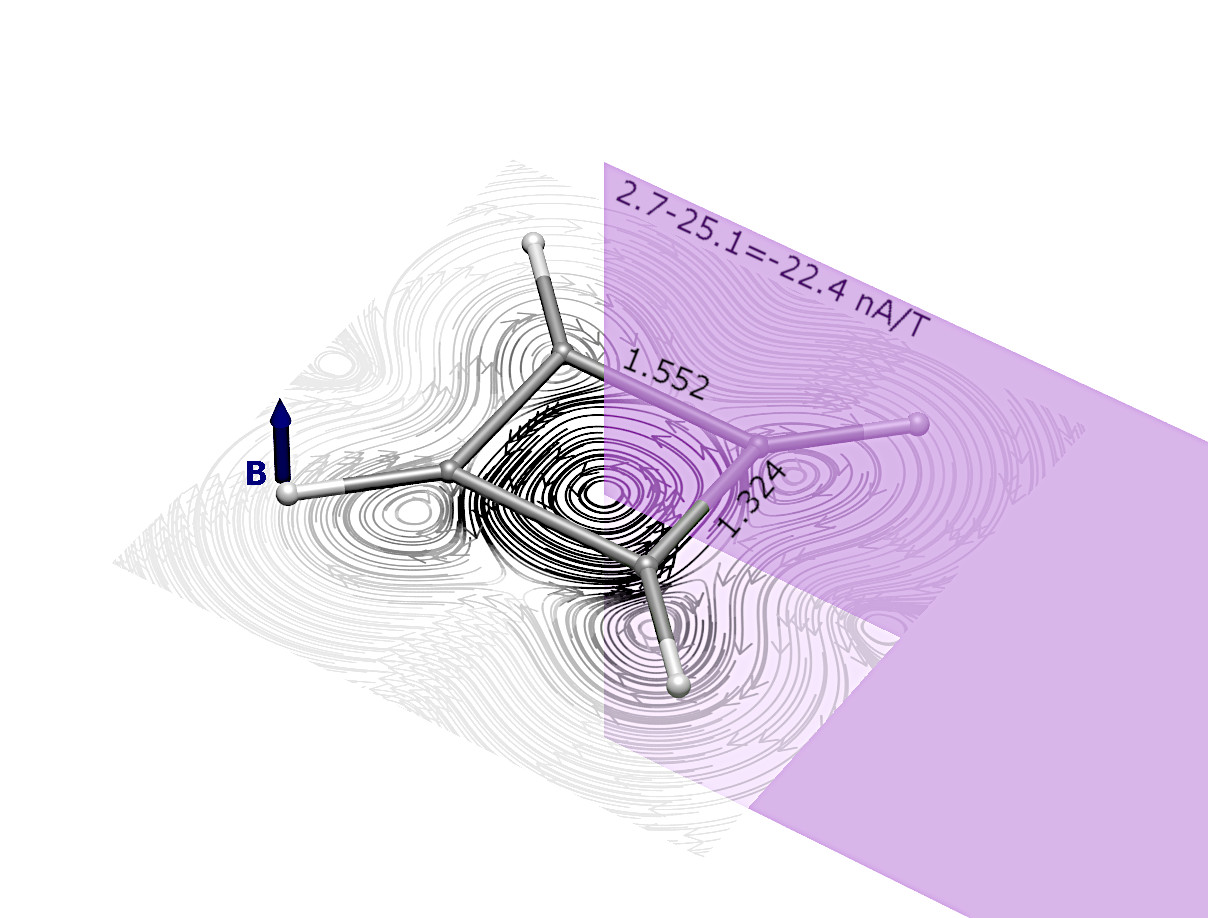}
\caption{$D_{2h}$ electronic ground state, minimum structure (distances given in units of \AA) of C$_4$H$_4$, and stream line plot of magnetically induced
 (field direction indicated by the blue arrow) currents at a plane 0.5 \AA, below the molecular plane. Integration of currents flowing through the 
purple surface yields a global molecular ring current of -22.4 nA/T (paratropic).\label{C4H4_currents}}
\end{figure}

\subsubsection*{C$_8$H$_8$}
Another example from the compound class of the $n$-annulenes is cyclooctatetraene (C$_8$H$_8$) 
which is similar to C$_4$H$_4$  in that it represents also a primoid second order JT case, but 
as we will see shows some quantitative difference in its paramagnetic response.\\ 

\begin{figure}
\centering
\tikzset{every picture/.style={line width=0.75pt}} 
\begin{tikzpicture}[x=0.75pt,y=0.75pt,yscale=-1,xscale=1]
\draw    (41.17,166.67) -- (292.17,169.31) ;
\draw [shift={(294.17,169.33)}, rotate = 180.6] [color={rgb, 255:red, 0; green, 0; blue, 0 }  ][line width=0.75]    (10.93,-3.29) .. controls (6.95,-1.4) and (3.31,-0.3) .. (0,0) .. controls (3.31,0.3) and (6.95,1.4) .. (10.93,3.29)   ;
\draw    (51,178) -- (50.01,31) ;
\draw [shift={(50,29)}, rotate = 449.62] [color={rgb, 255:red, 0; green, 0; blue, 0 }  ][line width=0.75]    (10.93,-3.29) .. controls (6.95,-1.4) and (3.31,-0.3) .. (0,0) .. controls (3.31,0.3) and (6.95,1.4) .. (10.93,3.29)   ;
\draw    (100,160) -- (100,177.67) ;
\draw    (169,160) -- (169,177.67) ;
\draw    (109,76.67) -- (89,76.67) ;
\draw    (178,134.67) -- (158,134.67) ;
\draw    (95.5,81.75) -- (95.07,69) ;
\draw [shift={(95,67)}, rotate = 448.06] [color={rgb, 255:red, 0; green, 0; blue, 0 }  ][line width=0.75]    (10.93,-3.29) .. controls (6.95,-1.4) and (3.31,-0.3) .. (0,0) .. controls (3.31,0.3) and (6.95,1.4) .. (10.93,3.29)   ;
\draw    (165.5,140.75) -- (165.07,128) ;
\draw [shift={(165,126)}, rotate = 448.06] [color={rgb, 255:red, 0; green, 0; blue, 0 }  ][line width=0.75]    (10.93,-3.29) .. controls (6.95,-1.4) and (3.31,-0.3) .. (0,0) .. controls (3.31,0.3) and (6.95,1.4) .. (10.93,3.29)   ;
\draw  [dash pattern={on 4.5pt off 4.5pt}]  (109,76.67) -- (158,134.67) ;
\draw    (102.4,67) -- (102.05,80) ;
\draw [shift={(102,82)}, rotate = 271.53] [color={rgb, 255:red, 0; green, 0; blue, 0 }  ][line width=0.75]    (10.93,-3.29) .. controls (6.95,-1.4) and (3.31,-0.3) .. (0,0) .. controls (3.31,0.3) and (6.95,1.4) .. (10.93,3.29)   ;
\draw    (172.4,126) -- (172.05,139) ;
\draw [shift={(172,141)}, rotate = 271.53] [color={rgb, 255:red, 0; green, 0; blue, 0 }  ][line width=0.75]    (10.93,-3.29) .. controls (6.95,-1.4) and (3.31,-0.3) .. (0,0) .. controls (3.31,0.3) and (6.95,1.4) .. (10.93,3.29)   ;
\draw    (109,74.67) -- (89,74.67) ;
\draw    (59.17,60.33) -- (42,60.67) ;
\draw    (240,160) -- (240,177.67) ;
\draw    (178,60.67) -- (158,60.67) ;
\draw    (250,155.67) -- (230,155.67) ;
\draw    (237.5,161.75) -- (237.07,149) ;
\draw [shift={(237,147)}, rotate = 448.06] [color={rgb, 255:red, 0; green, 0; blue, 0 }  ][line width=0.75]    (10.93,-3.29) .. controls (6.95,-1.4) and (3.31,-0.3) .. (0,0) .. controls (3.31,0.3) and (6.95,1.4) .. (10.93,3.29)   ;
\draw  [dash pattern={on 4.5pt off 4.5pt}]  (178,60.67) -- (230,60.67) ;
\draw    (244.4,147) -- (244.05,160) ;
\draw [shift={(244,162)}, rotate = 271.53] [color={rgb, 255:red, 0; green, 0; blue, 0 }  ][line width=0.75]    (10.93,-3.29) .. controls (6.95,-1.4) and (3.31,-0.3) .. (0,0) .. controls (3.31,0.3) and (6.95,1.4) .. (10.93,3.29)   ;
\draw    (250,60.67) -- (230,60.67) ;
\draw  [dash pattern={on 4.5pt off 4.5pt}]  (178,134.67) -- (230,155.67) ;
\draw  [dash pattern={on 4.5pt off 4.5pt}]  (109,74.67) -- (158,60.67) ;
\draw    (59.17,159.33) -- (42,159.67) ;
\draw (100,188) node  [align=left] {\textit{D}$\displaystyle _{8h}$};
\draw (168,188) node  [align=left] {\textit{D}$\displaystyle _{4h}$};
\draw (168.67,206.33) node  [align=left] {\textit{Q / arb. units}};
\draw (77,76) node   {$e_{2u}$};
\draw (28.67,60.33) node  [align=left] {0};
\draw (26.67,160.33) node  [align=left] {\mbox{-}10};
\draw (240,188) node  [align=left] {\textit{D}$\displaystyle _{2d}$};
\draw (146,136) node   {$b_{1u}$};
\draw (143,48) node   {$b_{2u}$};
\draw (260,157) node   {$a_{1}$};
\draw (260,62) node   {$a_{2}$};
\draw (19.67,99.33) node [rotate=-270] [align=left] {$\displaystyle \epsilon $\textit{ / }eV};
\end{tikzpicture}
\caption{Frontier orbital energies $\varepsilon$ for the ground state saddle point ($D_{8h}$ of second order, $D_{4h}$ of first order) or  
minimum structures ($D_{2d}$) calculated at the generalized valence bond - perfect pairing (GVB(PP)) level of theory for 
cyclooctatetraene (C$_8$H$_8$). Orbital energy splitting upon distortion $Q$ from $D_{8h}$ via $D_{4h}$ to the $D_{2d}$ minimum is indicated 
by the dashed lines. The first distortion $D_{8h} \to D_{4h}$ is a {\em primoid} second order Jahn-Teller (JT) distortion via a $B_{1g}$ mode, 
the second one is a non-primoid JT distortion via a $B_{1u}$ mode. Again, like in C$_4$H$_4$ the electronic ground state 
is non-degenerate (second order JT) but the underlying orbital levels show degeneracy and branching 
(primoid second order JT system). All three direct products $E_{2u}\otimes E_{2u}$, $B_{1u}\otimes B_{2u}$,
and $A_{1}\otimes A_{2}$, the latter two corresponding to the paramagnetic virtual excitation, contain or are 
identical to the IR for the $z$-component of the angular momentum operator. Virtual excitations represented by the terms 
given in expression \ref{c2} are not only decreasing by virtue of the increasing energy denominator but also due to the 
decreasing overlap integral in the numerator. Thus the molecule in the $D_{4h}$ saddle point configuration 
shows a much stronger paramagnetic response than the $D_{2d}$ minimum structure 
(compare Fig. \ref{C8H8_currents}).\label{C8H8_levels}}
\end{figure}
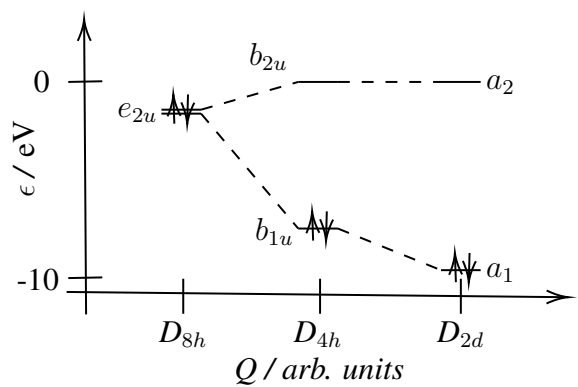

The idealized $D_{8h}$ symmetry is not observed as a ground state minimum\cite{10.1021/jp8067365} 
due to a primoid second order JT distortion ($e_u^2$ case). Also here we observe a $B_{1g}$ distortion
to a $D_{4h}$ structure, which turns out to be no minimum structure either as it undergoes another 
second order JT distortion along a $B_{1u}$ mode to a $D_{2d}$ symmetric minimum structure. 
As the first distortion is of primoid type we know that in $D_{4h}$, the IR $\Gamma_{\mathbf{l}_z}$ will 
be the tensor product of the branches of the $e_u$ orbitals. As these are $b_{1u}$ and $b_{2u}$ and 
$b_{1u} \otimes b_{2u} = a_{2g}$ and $\Gamma_{\mathbf{l}_z}=a_{2g}$ in $D_{4h}$ we find our symmetry 
rule confirmed. Since C$_8$H$_8$ undergoes another second order JT distortion into $D_{2d}$ we still find 
\begin{align}
(B_{1u} \otimes B_{2u})|_{D_{2d}} & =   A_{2g}|_{D_{2d}} \nonumber \\
B_{1u}|_{D_{2d}} \otimes B_{2u}|_{D_{2d}} & =   A_2 \nonumber \\
A_1 \otimes A_2 & =   A_2\nonumber 
\end{align}
In agreement with this $A_2$ represents $\Gamma_{\mathbf{l}_z}$ in $D_{2d}$, hence the prediction 
that the frontier orbital symmetries that were branching out of $e_u$ in this sequence of primoid 
and non-primoid second order JT distortions relate to each other in $\Gamma_{\mathbf{l}_z}$ 
manner is confirmed (again note that the electronic levels themselves are not degenerate and branch but 
rather couple by the distortion, see also Fig. \ref{C8H8_levels}). In addition paramagnetic response for the 
$D_{2d}$ minimum of C$_8$H$_8$ is predicted, since we started from a non-isometric 
point group symmetry.\\

\begin{figure}[H]
\centering
\begin{subfigure}{0.9\textwidth}
\centering
\includegraphics[width=0.8\textwidth]{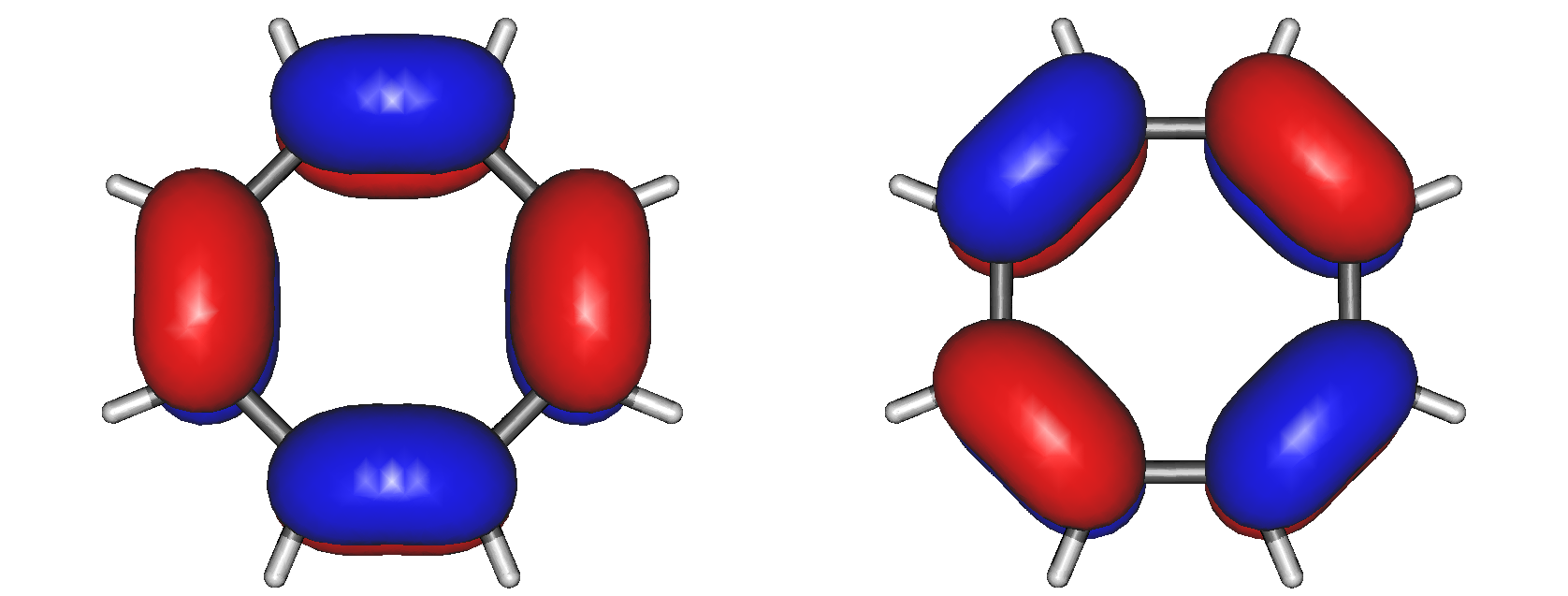}
\caption{Highest occupied molecular orbital (HOMO) in the $D_{8h}$ double saddle point structure. The two-fold degenerate orbital is represented by the $e_{2u}$ IR.}
\end{subfigure}
\begin{subfigure}{0.9\textwidth}
\centering
\includegraphics[width=0.8\textwidth]{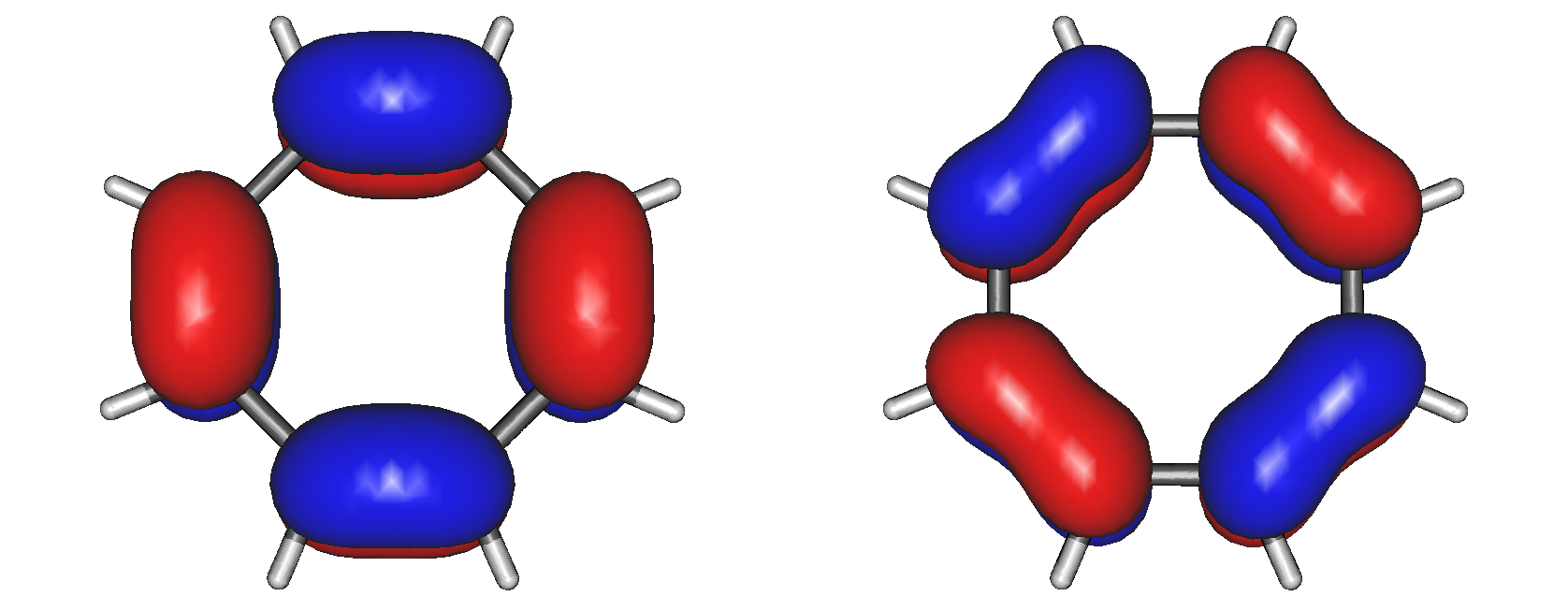}
\caption{HOMO (left) and lowest unoccupied molecular orbital (LUMO, right) in the $D_{4h}$ double saddle point structure. The HOMO is represented by $b_{1u}$, the LUMO by the $b_{2u}$ IR.}
\end{subfigure}
\begin{subfigure}{0.9\textwidth}
\centering
\includegraphics[width=0.8\textwidth]{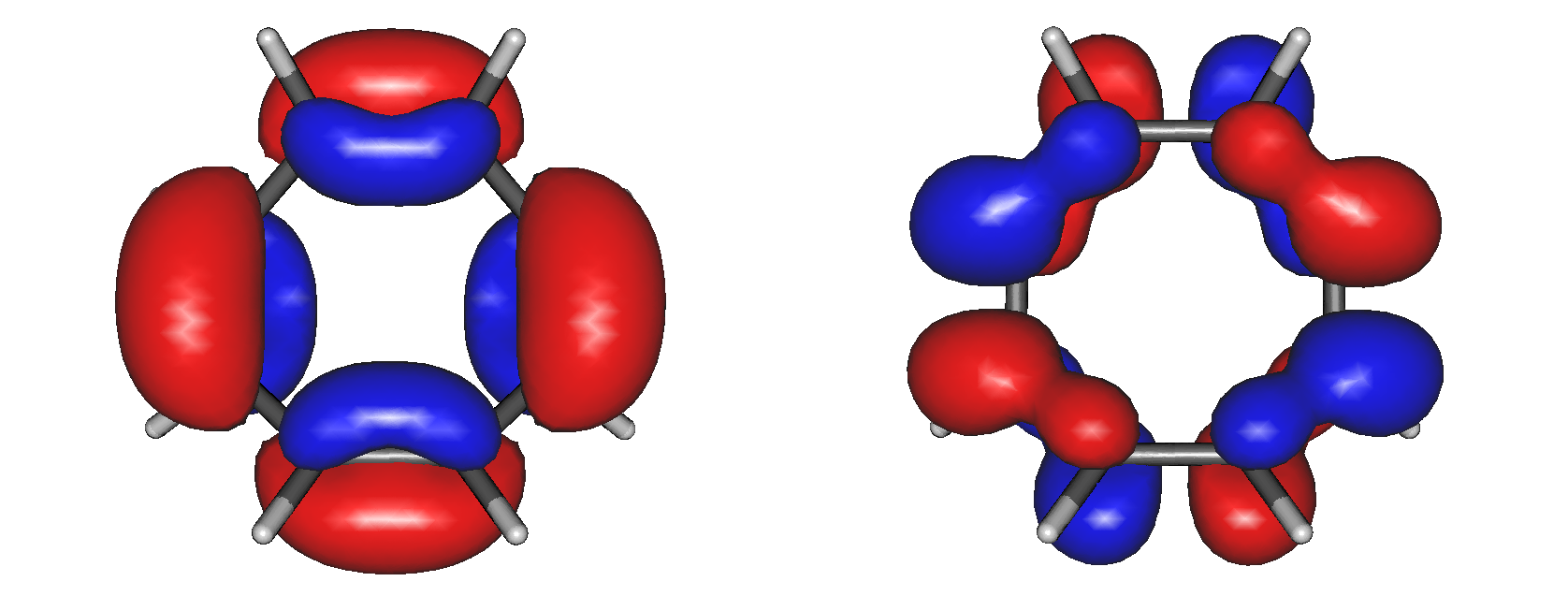}
\caption{HOMO ($a_1$, left) and LUMO ($a_2$, right) in the $D_{2d}$ minimum structure.}
\end{subfigure}
\caption{Frontier orbitals of C$_8$H$_8$ at different geometries, calculated at the GVB(PP)/TZVPP level of theory, displayed as 
isosurfaces at densities of 0.05.\label{C8H8-orbitals}}
\end{figure}

\begin{figure}[H]
\centering
\includegraphics[width=16cm]{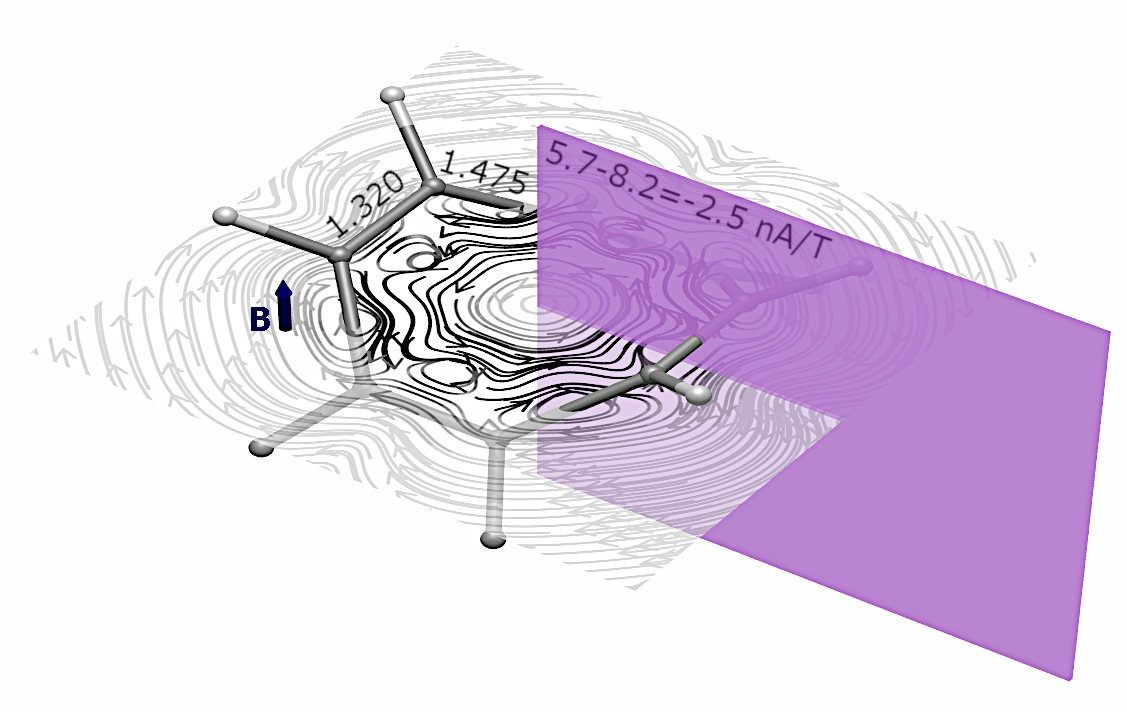}
\caption{$D_{2d}$ electronic ground state, minimum structure (distances given in units of \AA) of C$_8$H$_8$, and stream line plot of magnetically induced
 (field direction indicated by the blue arrow) currents in the average molecular plane ($x,y$-plane). Integration of currents flowing through the 
purple surface yields a global molecular ring current of -2.5 nA/T (purely paratropic contribution), which is comparably weak, thus the $D_{2d}$ electronic ground state minimum structure of C$_8$H$_8$ is classified as weakly antiaromatic or non-aromatic. \label{C8H8_currents}}
\end{figure}

In contrast to the markedly paramagnetic nature of the magnetic response of the ground state minimum of 
C$_4$H$_4$ (-21 nA/T) for C$_8$H$_8$ we find barely
-2.5 nA/T total induced current. However, the paratropic contribution amounts to -8.2 nA/T, thus we again see the rule 
confirmed that first and primoid second order JT distortions lead to systems with paramagnetic response.
In this case however the total paramagnetic response is too weak to assign the predicate ``antiaromatic'' to this molecule. 
One reason is that the {\em second} second order JT distortion (which for itself is not of primoid type since the degeneracy 
of the frontier orbital has already been lifted to $b_{1u}$ and $b_{2u}$) causes a further increase of the HOMO-LUMO 
energy gap to 9.7 eV at GVB(PP) level of theory, thus diminishing the magnetic response accordingly. For 
comparison the value for C$_4$H$_4$ at the same level of theory (GVB(PP)) is 7.2 eV.

\subsubsection*{The General 4$n$ $\pi$ annulene and double bond localization (bond length alternation)}

In antiaromatic $n$-annulenes (C$_{4n}$H$_{4n}$) in the idealized $D_{(4 n)h}$ symmetry, 
double bond localisation can be easily understood as a second order (primoid) JT effect. The saddle point in
these cases would have a double occupied  $e_{nu}$ or $e_{ng}$ orbital, for even $n$ or for odd $n$, respectively, 
thus an ($e_{nu}$)$^2$ or ($e_{ng}$)$^2$ configuration, respectively, that can distort via a $b_{1g}$ mode to 
$D_{(2 n)h}$ symmetry, leading to a branching of $e_{nu}$ to $b_{1u}$ and $b_{2u}$ 
for even $n$ and similarly with {\em gerade} symmetry species $_g$ for odd $n$, 
which would correspond in analogy to the above discussed C$_8$H$_8$ to a double bond localization in the 
first instance.

An introductory discussion of the view of double bond localisation in ring systems as a JT distortion 
as well as its relation to the Peierls distortion in polyethin is given in ref. \cite{Hoffmann}.

\subsubsection*{MnF$_3$}

\begin{figure}[H]
\centering
\includegraphics[width=0.8\textwidth]{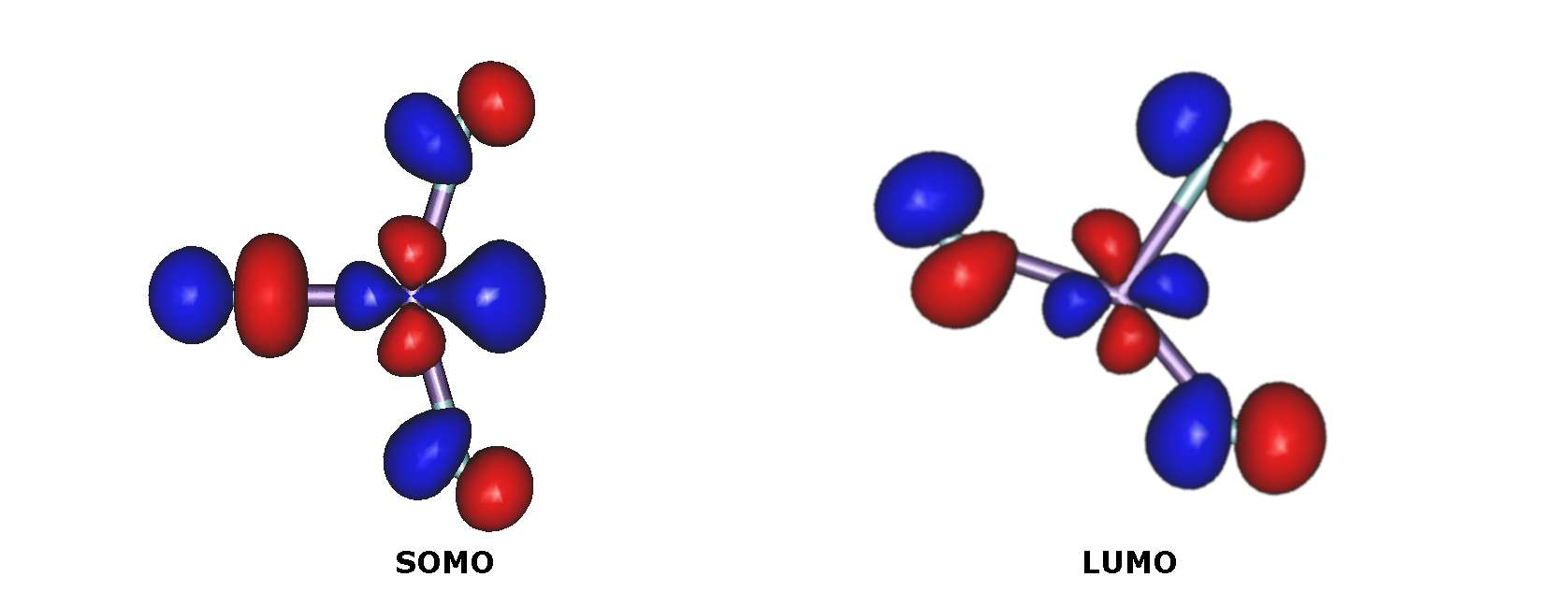}
\caption{The $^5B_1$ electronic ground state structure of MnF$_3$ with $C_{2v}$ symmetry and 
the $b_1$ $\alpha$-HOMO and $a_1$ $\alpha$-LUMO. The LUMO is displayed for a rotated molecule, the angle of rotation is about -120°. 
The rotational relation between the orbitals is already noticeable with bare eye.\label{MnF3_orbs}}
\end{figure}

\begin{figure}[H]
\centering
\includegraphics[width=0.8\textwidth]{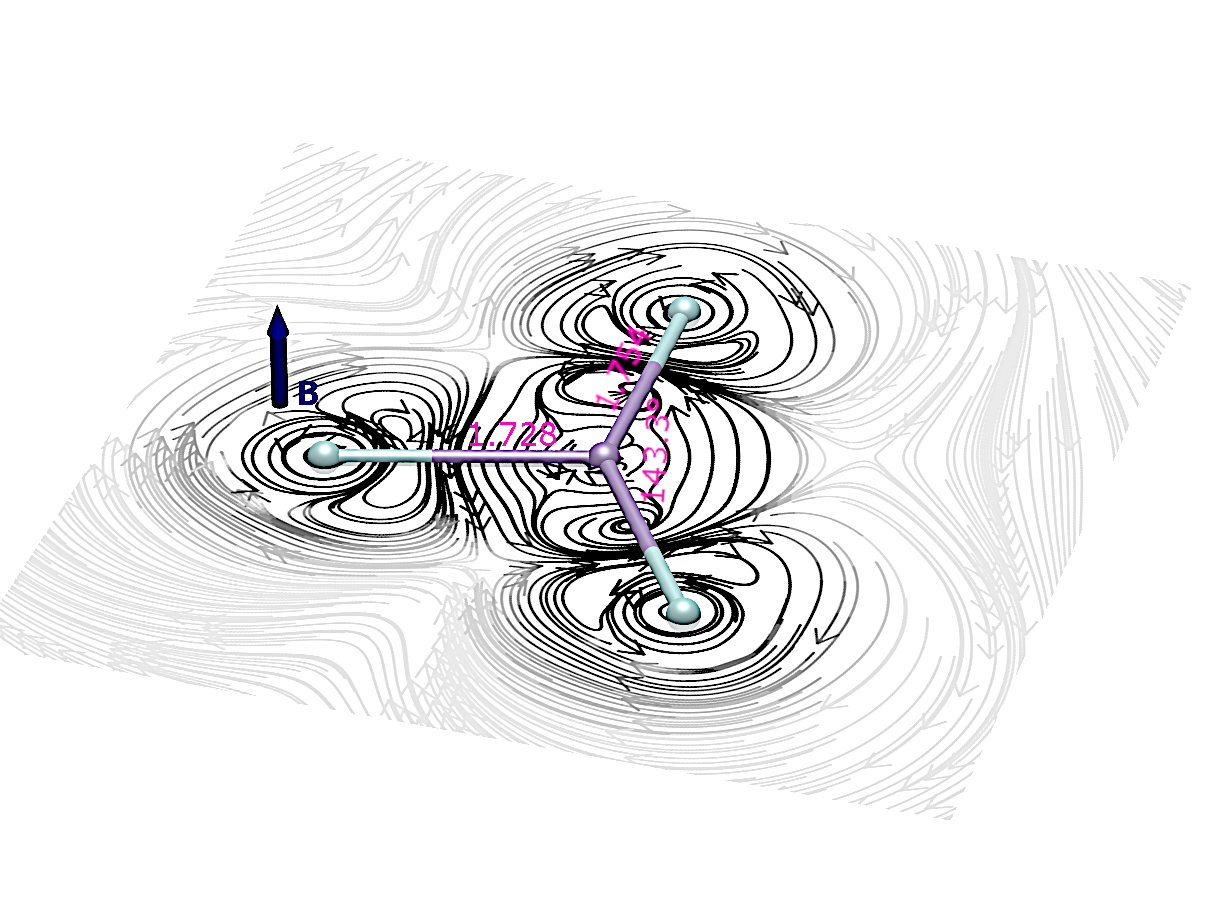}
\caption{The $^5B_1$ electronic ground state structure of MnF$_3$ with $C_{2v}$ symmetry and a stream-line plot of the
magnetically induced currents in the molecular plane ($\mathbf{B}$ field is set perpendicular to the molecular plane).
A paratropic current vortex (counterclockwise) around the Mn atom is found.\label{MnF3_currents}}
\end{figure}

Manganese trifluoride is a prime example for a first order Jahn-Teller effect which would give in $D_{3h}$ 
(a non-isometric point group) an $E'$ ground state. In fact the lowest energy species is in the $^5B_1$ 
electronic ground state and has a planar $C_{2v}$ symmetric structure which was predicted theoretically and confirmed 
by gas-phase electron-diffraction \cite{10.1021/ja9712128}. 
Magnetic response calculations indeed show magnetically induced 
paratropic currents around the central manganese atom as well as locally around the fluorine atoms 
(see Figures \ref{MnF3_orbs} and \ref{MnF3_currents}).\\

Unlike in the previous two examples the induced paramagnetism is hidden to some extent, 
since experimentally the permanent magnetic moment resulting from spin dominates the magnetic properties 
and for the other since in this case the paramagnetic currents are atom-centered which hampers 
the computational analysis of the current densities due to interference with other
atomic current contributions. In addition though, 
MnF$_3$ can be observed in the gas-phase\cite{10.1021/ja9712128}, in condensed 
matter the coordinatively unsaturated threefold coordinated
Mn$^{3+}$ cation under normal conditions is not observable. 

\subsubsection*{Pentalene}

An interesting case that at first sight seems to be an outlier of our antiaromaticity symmetry rule is pentalene
(see below for a molecular formula). Pentalene is a well-known case of antiaromaticity and its magnetisability was 
recently studied computationally \cite{10.1039/c6cp01968d}. Its apparent JT idealized symmetry 
would be $D_{2h}$ (with HOMO $a_{u}$ and LUMO $b_{1u}$) while its computationally predicted and experimentally 
confirmed minimum structure is $C_{2h}$ (HOMO: $a_{u}$, LUMO $a_{u}$). 
However, $D_{2h}$ is clearly a class $\mathcal{K}_{aaa}$ point group, thus the observed $D_{2h} \to C_{2h}$ 
distortion cannot be interpreted as a primoid second order JT type distortion (as outlined in section \ref{non-primoid}), 
nevertheless we observe strong paramagnetic response which is dominated by the HOMO-LUMO virtual excitation and a 
distortion that is in agreement with a double-bond localisation. In essence the crucial argument in this case 
is that we have to deal with a ring system containing a set of conjugated double bonds on its perimeter. Thus a 
natural way of interpretation would be to classify pentalene as a perturbed eight membered ring, that 
derives from an idealized but perturbed D$_{8h}$ point group symmetry, with 
the perturbation being the additional transannular C-C bond. In this way the fact that the direct products 
of the HOMO-LUMO IRs are identical to the IR of $\mathbf{l}_z$ is mostly easily seen from the branching of 
$D_{8h}$'s $E_{2u}$ into the $A_u$ and $B_{1u}$ IRs.

\begin{figure}[H]
\centering
\includegraphics[width=3.5cm]{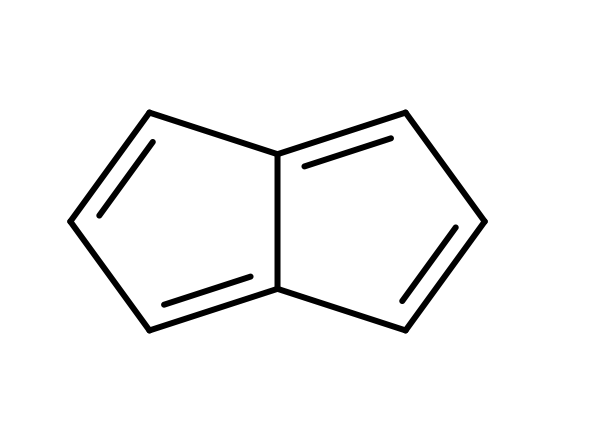}
\caption*{pentalene}
\end{figure}

Therefore pentalene cannot directly by using our symmetry rule be classified as an 
antiaromatic compound.

\subsection*{Chemical interpretation of dia- and paratropic response currents}

Equation \ref{L} means that a magnetic field probes the accessibility of virtual states that 
ideally differ from the ground state only by a rotational orientation of the wave function 
in space and thus are spatially very closely related. In case a molecule responds with a 
strong paratropic current density this can mean that there is a situation of close to 
degeneracy resembling the spatial symmetry of an atomic open subshell. The optimal 
symmetry/energy relation in question of eq. \ref{L} in this way can 
be interpreted as the analog to an atomic open subshell case for a general molecular system, and 
the magnetic response is a concise probe for that. In addition one also yields information on the 
direction of the symmetry relation from the direction of the field $\mathbf{B}$.\\

The type of degeneracy that is probed by the magnetic field is 
not the only possible type. Other types of (near) degeneracy include ones derived form what 
is sometimes called ``accidental'' degeneracy where there is no obvious (close-to-)symmetry 
relation in $O(3)$ between occupied and empty levels and hence no contribution to 
paramagnetic response from these. \\



In contrast the antithetical concept of aromaticity is characterized by
induced diamagnetic currents. In the common gauge origin 
approach, diamagnetic currents to first oder are determined 
by the ground eigenstate $\Psi_0$ of the $\mathbf{B}$ field free system and 
correspond to a precession motion of the electrons 
in the magnetic field that increases the average expected angular momentum 
$\braket{\Psi|\hat{l}_z|\Psi}$, (for $\Psi$ the wave function in the magnetic field) 
which will lead in general if the ring current extends over a sufficiently large
domain covering sufficiently much of the total electron density in the molecule 
to an increase in total energy.\\

An alternative way is to describe diamagnetic response via the
``Continuous Transformation of the Origin of the Current Density'' (CTOCD-DZ) 
approach that was introduced to theoretical chemistry by Keith \& Bader
in 1992\cite{10.1016/0009-2614(93)89127-4}  that allows in particular within 
to Fowler \& Steiner's ipsocentric {\em ansatz}\cite{10.1039/B104847N} to 
express the full magnetic response, completely in terms of occupied to virtual state (or orbital) 
transitions.\\

In this model the diamagnetic currents are determined by 
virtual transitions of dipole symmetry, namely dipole transition
moments perpendicular to the external field $\mathbf{B}$. That means 
that for example sigma bonds that leave pairs of $\sigma$ and $\sigma^*$ type occupied 
and virtual orbitals, where the bond direction is perpendicular to the external field,
contribute to the diamagnetic response. A simple example would be a bonding orbital 
that has some anti-bonding counterpart with a nodal plane parallel to $\mathbf{B}$.\\

In that way an electronic structure that contains covalent or systems of 
delocalized covalent bonds (and naturally the corresponding anti-bonding orbitals
unoccupied) is diamagnetic. So we can summarize that a strong diamagnetic response current 
can be a consequence of (full) occupation of bonding orbitals and empty anti-bonding orbitals,
which differ from bonding orbitals by additional nodal planes parallel to $\mathbf{B}$. 
Whereas paramagnetic response can indicate incompletely occupied molecular shells which are 
similar up to a rotation around an axis parallel to  $\mathbf{B}$. This in effect leads to 
a distinctly different picture of antiaromaticity as it was originally envisaged by Breslow 
as he for instance noted ``Thus antiaromaticity is a particular aspect of anti-bonding, 
just as aromaticity is a particular aspect of bonding''\cite{doi:10.1021/ar50072a001}, which 
is according to our analyses not the case since its not the occupation of
anti-bonding orbitals that lead to the characteristic paramagnetic response currents, 
nor would it lead to a peculiar small HOMO-LUMO energy gap in general, 
or be connected with structural distortions.
\\

At quick glance it might be surprising that antiaromaticity has often been observed or at least proposed 
for organic ring systems and related species like polycyclic molecules. In contrast antiaromaticity 
is rarely mentioned in the context of inorganic coordination compounds and metal clusters. On 
the bases of our findings this is easily explained, because the vast majority of organic 
ring systems and related compounds can be deduced from the particular point groups $\mathcal{K}_{ae}$, that 
is non-isometric point groups with degenerate real irreducible representations. In exactly this 
symmetry family we have the strict relation between JT distortion and paramagnetic response. While the majority of
3D-cluster systems or coordination compounds are derived from isometric point groups where this 
strict correspondence does not exist. In addition coordination compounds are of course metal-centered
and at least experimentally the induced paramagnetism there is often masked by permanent magnetic moments
centered at the metal atoms.\\

The here introduced analyses might serve as a basis for a deeper understanding of antiaromaticity, 
but despite its principle significance, antiaromaticity is not a widely spread chemical phenomenon
(for obvious reasons, since it is not only connected with structural instability but also instability 
in the sense of high reactivity by virtue of the small energy gap between occupied and virtual states).
At this stage one might speculate if there is some consistent way to define aromaticity by the absence of 
paramagneticity alone, an idea going back to Bilde \& Hansen\cite{10.1080/002689797170437}. 
In that way the presence or absence of rotationally accessible virtual states and their intimate 
connection with structural distortions could eventually be the only decisive criteria for presence 
or absence of aromaticity or antiaromaticity.\\

An extension of these considerations we are currently working on is the analysis of the symmetry 
rules of magnetic response and JT distortion in the fully relativistic domain including spin-orbit coupling.

\section*{Conclusion}

We have shown by symmetry considerations how Jahn-Teller distortion and paramagnetic response are intimately 
related in the family of points groups comprised of $C_n$, $C_{nv}$, $C_{nh}$, $D_n$, 
$D_{nh}$, for $n>2$ and $ S_{2n}$ and $D_{nd}$ for $n>1$. We suggest that this connection 
is the underlying symmetry principle of
antiaromaticity, since the latter is defined by the central properties of a small HOMO-LUMO gap, proneness to 
structural distortion or instability and magnetically induced paramagnetic ring currents. We also saw that 
electronic structure related to covalent or delocalized covalent bonding or filled atomic subshells 
followed by empty subshells is related to diamagnetic response while electronic situations resembling 
open atomic subshells are related to paramagnetic response. The sparsity of examples may not only go back 
to the fact that antiaromaticity is related to low energetic stability but also due the paradoxy 
between distortion and paramagneticity that is hidden under its definition and 
that we have revealed in this work.

\bibliographystyle{naturemag}
\bibliography{scibib}

\section*{Acknowledgments}

The work has been performed under the Project HPC-EUROPA3 (INFRAIA-2016-1-730897), with the support of the EC Research Innovation
Action under the H2020 Programme; in particular, RB gratefully acknowledges the support of Prof. Dr. Dage Sundholm from 
the University of Helsinki
and the computer resources and technical support provided by the CSC Helsinki. RB also acknowledges Prof. Dr. Dage Sundholm 
for financial support and Prof. Dr. Martin Breza from the Slovak University of Technology 
(Bratislava), Prof. Dr. Paolo Lazzeretti (Modena) and Prof. Dr. Juha Vaara (Oulu) for fruitful discussions 
and Prof. Dr. Nicola H\"using (Salzburg) for generous and kind support of all research activities. 

\section*{Author Contributions}

RB provided the original idea performed the computational work and wrote the manuscript. AV provided the mathematical proof
and the wrote the respective parts in the manuscripts.

\section*{Competing Interests}

We declare that there were no competing interests. 

\newpage
\begin{center}
\textbf{\Large Supplementary Information}\\
\textbf{\large The principle underlying antiaromaticity}\\
\textbf{Raphael J. F. Berger$^{1\ast}$ and Alexandre Viel$^2$}
\\
\normalsize{$^{1}$Department for Chemistry and Physics of Materials, University of Salzburg,}\\
\normalsize{Jakob-Haringerstr 2a, A-5020 Salzburg, AUSTRIA}\\
\normalsize{$^{2}$8 Sente de la haie Saint-Marc, 78480 Verneuil sur Seine, FRANCE}\\
\normalsize{$^\ast$To whom correspondence should be addressed; E-mail:  raphael.berger@sbg.ac.at.}
\end{center}
\vspace{3cm}
\setcounter{equation}{0}
\setcounter{figure}{0}
\setcounter{table}{0}
\setcounter{page}{1}
\setcounter{section}{0}

\makeatletter
\renewcommand{\theequation}{S\arabic{equation}}
\renewcommand{\thefigure}{S\arabic{figure}}
\renewcommand{\thepage}{S\arabic{page}}
\renewcommand{\thesection}{S\arabic{section}}
\renewcommand{\bibnumfmt}[1]{[S#1]}
\renewcommand{\citenumfont}[1]{S#1}

\beginsupplement

\section{Group and representation theoretical details}

\label{S1}
While in mathematics a representation is ``of'' a group and ``acting on a vector space'', placing the group first in order of importance, 
we instead talk in the main part this work of ``representations of vector spaces'' or ``of operators'', ``by a group''. 
This different phrasing is because the operators are what is more concrete in chemistry than the abstract groups.\\

A {\em point group} is a subgroup of $O(3)$, or from a more mathematical point of view, 
an abstract group together with a faithful representation of itself into $O(3)$. 
Thus two point groups can be isomorphic as abstract mathematical groups but be different representations 
at the same time (e.g. $C_s$ and $C_2$), and for that be different point groups.
The reason we want to separate the discussion of the group from the discussion of its representation is that
we naturally encounter several representations by the same group.\\ 
 
The physical space ${\Bbb R}^3$ upon which the group elements act, serves as the representation space implied
by the concept of a ``point group''. Hence its axis ($x,y,z$) represent the algebraic generators of the 
physical quantities corresponding to operators or observables, which thereby can be analysed in 
terms of irreducible representations.\\

Once we have this representation $\rho$ of physical space, we often want to describe
its action on algebraically related spaces, such as direct sums, tensor spaces or dual spaces, 
or in the present case, angular-momentum space.
Computing the representations of those more complex spaces from the initial representation 
of physical space can be described mathematically with the use of Schur functors. 
The functor that will interest us in order to describe the representation of the angular momentum 
operator is the second exterior power :\\

If we have a change of variable represented by an orthogonal linear map $\mathbf r' = U\mathbf r$, then 
$\nabla' = U \nabla$ because $U$ is orthogonal, so that when transforming the angular momentum operator,
we get $\mathbf{l}' =  -i\hbar({\mathbf r'} \times \nabla') = -i\hbar(U{\mathbf r} \times U\nabla) = (U \times U) \mathbf{l}$
where $U \times U$ (which can also be written $U \wedge U$, $[U \otimes U]$ or $Alt^2(U)$) is the application of 
the second exterior power to $U$.
Finally since a representation is nothing but a family of orthogonal linear maps, we can also apply Schur functors to representations,
so that if $\rho$ is a representation of the physical space, $Alt^2(\rho)$ is the representation of the angular momentum operator.\\

Once we have identified this relationship, we gain the advantage of having computation rules,
such as $Alt^2(\rho_1 \oplus \rho_2) = Alt^2(\rho_1) \oplus Alt^2(\rho_2) \oplus \rho_1 \otimes \rho_2$.
The symmetric square functor $Sym^2$ is also of interest, it satisfies the same rule, and we have the functor decomposition
$\rho \otimes \rho = Sym^2(\rho) \oplus Alt^2(\rho)$.
Finally, if $\Gamma$ is a real irreducible representation by a group $G$, then $Sym^2(\Gamma)$ always has exactly one 
totally symmetric $\Gamma_0$ component in its decomposition, so that for any two dimensional real irreducible representation $E$,
\begin{equation}E\otimes E = Alt^2(E) \oplus \Gamma_0 \oplus \rho_q;\;\; dim(\Gamma_0) = 1,\, dim(\rho_q)=2. \label{E2_in_G}\end{equation}\\ 

We also make a note that functors commute with restrictions. If $F$ is a functor, $\rho$ a representation by a group $G$, and $H$
a subgroup of $G$, then for any $h \in H$, \\
$F(\rho|_H)(h) = F(\rho|_H(h)) = F(\rho(h)) = (F(\rho))(h) = (F(\rho))|_H(h)$,
so that $F(\rho|_H) = (F(\rho))|_H$. This works exactly the same way for bifunctors such as direct sums or 
tensor products of two representations. \\

We investigate what this entails when we look at the representation of the angular momentum operator of point groups.
The main result we will want to prove is that the $Alt^2$ functor has a tendency to output the same
(irreducible) representation when you give as input different irreducible representations of the same dimension.\\

All point groups can be partitioned into three families:
\begin{enumerate}

\item Point groups whose faithful representations decompose into three components, so that they 
      embed in $O(1) \times O(1) \times O(1)$, {\em i.e.} groups that have no axis of rotation of order 
      $n>2$ (family $\mathcal{K}_{aaa}$). In detail $\mathcal{K}_{aaa}=\{C_1$, $C_i$, $C_2$, $C_s$, 
      $C_{2v}$, $C_{2h}$, $D_2$, $D_{2h}\}$.
  
\item Point groups whose faithful representations decompose in two irreducible components but not three, so they 
      embed in $O(1) \times O(2)$, {\em i.e.} groups that have a unique main axis of rotation of highest order 
      $n$ (family $\mathcal{K}_{ae}$). In detail $\mathcal{K}_{ae}  =\{C_n$, $C_{nv}$, $C_{nh}$, $D_n$, 
      $D_{nh}$, $D_{(n-1)d}$, $S_{2(n-1)}\}$, for $n>2$

\item Point groups whose faithful representations are irreducible, {\em i.e.} groups that have more than one axis of 
      rotation of order $n>2$, which are also called isometric groups (family $\mathcal{K}_{t}$). 
      In detail $\mathcal{K}_{t} = \{T$, $T_d$, $T_h$, $O$,$ O_h$,$ I$,$ I_h\}$.\\

\end{enumerate}

It is not obvious that this partition 
defined in terms of a specific representation, carries down to the level of abstract groups 
independent of specific representations and we note that this is not true in general for 
abstract groups represented in $O(n)$ for $n>3$.\\

Now we consider a point group from the first two families, whose representation is reducible, such that it 
decomposes into a direct sum $\rho = A \oplus E$.\\ 

An equivalent property is that the representation has a unique main symmetry axis, and then this axis is the subspace acted on by $A$  :
Let $n$ be the maximum order of a rotation of the representation and consider the set of all the axis of the order $n$ rotations.
The group acts on this set of axis as the group acts by conjugation on the rotations :
if a line $l$ is the axis of a rotation $\rho(g)$ and $h \in G$, then $\rho(hgh^{-1})$ is a rotation whose axis is $\rho(h)(l)$.
If there is only one such axis $l$ (called the main symmetry axis), it means that $\rho(h)(l) = l$ for all $h \in G$, 
that is $l$ is a subspace stabilized by $G$, so $\rho$ is decomposes as
a direct sum of the action on $l$ and the action on its orthogonal. 
We will thus call this $l$ (or $A$'s underlying vector space) the $z$-axis and its orthogonal (or $E$'s vector space) the $(x,y)$-plane.\\

In this $\mathcal K_{ae}$ family, the angular momentum operator $Alt^2(\rho)$ decomposes into $Alt^2(\rho) = Alt^2(A) \oplus Alt^2(E) \oplus A \otimes E$.
Since $A$ is one dimensional, $Alt^2(A)$ is zero-dimensional and can be omitted :
$Alt^2(\rho) = Alt^2(E) \oplus A \otimes E$.

Thus for point groups from $\mathcal{K}_{ae}$ the (quantum mechanical) angular momentum (operator) decomposes into a one 
dimensional, hence irreducible, representation on the rotations around the $z$-axis $\Gamma_{\hat l_z} = Alt^2(E)$ (which
is also the anti-symmetric square of the representation of the $(x,y)$-plane) and
a two-dimensional representation $\Gamma_{\hat l_{x,y}} = A \otimes E$ on the rotations whose axis lie in the $(x,y)$-plane.\\

Moreover, $Sym^2(E)$ here further decomposes into $\Gamma_0 \oplus \rho_q$ where $\Gamma_0$ is the totally symmetric IR 
and $\rho_q$ is some two-dimensional representation. In summary for groups from  $\mathcal{K}_{ae}$ we have:

\begin{equation}
  E \otimes E = \underbrace{\Gamma_{\hat{l}_z}}_{ = Alt^2(E)} \oplus \underbrace{\Gamma_0 \oplus \rho_q}_{ = Sym^2(E)};\;\; \dim(\Gamma_0) = \dim(\Gamma_{\hat{l}_z})=1,  \dim(\rho_q)=2.
\label{E2_in_K_SI}
\end{equation}

We now want to prove the following property of their underlying abstract group, 
that given two (two-dimensional real irreducible) representations $E_1, E_2$ of point groups form $\mathcal{K}_{ae}$, 
then $Alt^2(E_1) = Alt^2(E_2)$, so that 

\begin{equation}
 \forall G \in\mathcal{K}_{ae}, \forall E \in Irrep(G,2),  Alt^2(E) = \Gamma_{\hat l z}  \label{lz_in_K}
\end{equation}

We start with finding out the possible underlying abstract groups from the families $\mathcal{K}_{aaa}$ 
and $\mathcal{K}_{ae}$ and show they belong to one of the four families 
$C_n,D_n,C_n\times C_2$ and $D_n \times C_2$. It is easy to see that $C_{nv}$ and $D_n$, $C_{nh}$ and  
$C_n \times C_2$, $D_{nh}$ and $D_n \times C_2$, $D_{nd}$ and $D_{2n}$, and $S_{2n}$ and $C_{2n}$ are isomorphic.
Then let $G$ be a finite group faithfully represented in $O(1) \times O(2)$.
This means we have two maps $\phi_i : G \to O(i)$ such that $G$ is a subgroup of $G' = \phi_1(G) \times \phi_2(G)$. 
If $\phi_2$ is not an isomorphism of $G$ to $\phi_2(G)$, then $|G'| \ge |G| > |\phi_2(G)|$.
But since $|G'|/|\phi_2(G)| = |\phi_1(G)|$ can only be $1$ or $2$, we must have $|\phi_1(G)|=2$ and $|G'| = |G|$, so that $G = G'$.
This proves that $G$ is isomorphic to either a finite subgroup of $O(2)$, or to a product of one with $C_2$ .\\

We now prove the property for the finite subgroups $C_n$ and $D_n$ of $O(2)$.
$C_n$ are abelian groups, so their complex irreducible characters are all one-dimensional. 
Given a choice of generator of $g \in C_n$, for each $n$th root of unity $\zeta$ there is one 
character $\chi_\zeta$ such that $\chi_\zeta(g) = \zeta$. The real irreducible representations
are obtained by pairing $\chi_\zeta$ with $\chi_{\overline \zeta}$ when $\zeta$ is not real and 
building their direct sum. Then, $Alt^2(\chi_\zeta \oplus \overline{\chi_\zeta}) = \chi_\zeta \otimes 
\overline{\chi_\zeta} = \chi_1 = 1$.\\

The dihedral groups $D_n$ are generated by two elements $r$ and $s$ subject 
to the relations $r^n=s^2 = rsrs = e$. They have complex two-dimensional 
irreducible representations $\rho_k$, given by the assignments 
$$r \mapsto \begin{pmatrix} \cos \frac{2k\pi}n & -\sin \frac{2k\pi}n\\\sin \frac{2k\pi}n &  \cos \frac{2k\pi}n\end{pmatrix},\;\mathrm{and}\; 
s \mapsto \begin{pmatrix} 1 & 0 \\ 0 & -1 \end{pmatrix},\; \mathrm{for} \; 0 < k < n/2.$$
Then $Alt^2(\rho_k)(r) = (1)$ and $Alt^2(\rho_k)(s) = (-1)$, which are independent of $k$.\\
This proves the claim for the $C_n$ and $D_n$ families. As for the $C_n \times C_2$ and $D_n \times C_2$ families, 
their irreducible representations are tensor products of one irreducible representation of each.
Since we are in dimension 2, for any $E$ representation of $C_n$ or $D_n$ and $A$ representation of $C_2$,
we have $Alt^2(E \otimes A) = Alt^2(E) \otimes (A \otimes A) = Alt^2(E) \otimes \Gamma_0$, which is independent
of the choice $E$ and $A$.\\

The determination of the irreducible representations of these groups shows that our 
classification of point groups is also a classification at the level of abstract groups:
we have shown in all cases that there was no three-dimensional irreducible representation of $G$.
So in particular the families $\mathcal{K}_{aaa}$ and $\mathcal{K}_{ae}$ of groups are disjoint 
from the family $\mathcal{K}_t$.

Moreover, if $n \ge 3$ then $G$ cannot be faithfully represented by a direct sum
of three one dimensional representations as then $G$ cannot have any element of 
order higher than two. Thus any faithful representation of these groups in $O(3)$ has 
an $A \oplus E$ shape and has a unique main axis of symmetry. These groups form 
the family $\mathcal{K}_{ae}$.

Finally, if $n \le 2$ then there are no two-dimensional irreducible representations of 
$G$, so these groups form the $\mathcal{K}_{aaa}$ family. In their case, the choice of 
a $z$-axis needs more than just algebraic considerations.\\

Consider now a point group from the family $\mathcal{K}_{t}$, {\em i.e.} an isometric point group, 
it comes with a faithful irreducible representation $T$ into $O(3)$. 
Such a point group cannot distinguish between the three coordinate axis $x$,$y$ and $z$.
Then the angular momentum representation $\Gamma_{\hat l} = Alt^2(T)$ is irreducible.
For three-dimensional representations, $Alt^2(Alt^2(T)) = \det(T) \otimes T$, so if $Alt^2(T)$ was reducible,
the reducibility would carry over to $\det(T) \otimes T$, and then 
by tensoring again with $\det(T)$, to $T$ itself.\\

Among the cubic point groups, this is a subfamily of the isometric point groups consisting of $T, T_d, T_h, O$ and $O_h$, a similar 
property is true: if $T_1$ and $T_2$ are two irreducible three-dimensional
representations, then $Alt^2(T_1) = Alt^2(T_2) = \Gamma_{\hat l}$. This is because in every case, there is a one dimensional representation $A$ 
such that $T_2 = A \otimes T_1$, so that $Alt^2(T_2) = Alt^2(T_1) \otimes A \otimes A = Alt^2(T_1)$. Among the icosahedral groups ($I, I_h$), 
this is no longer true. Here also the symmetric square $Sym^2(T)$ further decomposes into $\Gamma_0 \oplus \rho_q$ where $\rho_q$ is some five-dimensional representation.\\  
%
%

\section{Computational details}

\subsection{C$_4$H$_4$}

The geometry optimization of 1,3-cylcobutadiene in $D_{4h}$ symmetry 
was performed with the GAMESS-US\cite{10.1002/jcc.540141112} programme package 
(version {\verb 5 DEC 2014 (R1)}) for a reason lined out below and in the spirit of 
an early work from Voter and Goddard at the generalized valence bond - 
perfect pairing (GVB(PP)) level of theory\cite{10.1021/ja00271a008} and using the Karlsruhe def-TZVPP basis 
sets for carbon and hydrogen \cite{10.1063/1.463096}. Starting orbitals and geometry were used from a 
restricted Hartree-Fock (RHF) calculation for the dianionic species [C$_4$H$_4$]$^{2-}$, restricted to 
$D_{4h}$ spatial symmetry (but the wave function is effectively treated in GAMESS as $D_{2h}$ since no 
higher wave function symmetries are available in the current versions of GAMESS-US). For the sub-sequential
GVB run 13 doubly occupied inactive orbitals and 1 electron pair in 2 orbitals were set. That setting 
completely is equivalent to a complete active space two electrons in two orbitals computation (CAS(2i2)). We note that
convergence of GVB in cases can be considerably slower and in cases of problems with the starting orbitals 
it seems convenient to converge the orbitals in an actual CAS(2i2) run, that for example can be run using 
the full-Newton-Raphson converger that is implemented in GAMESS-US. The main reason for choosing GVB(PP) instead of 
CAS was to be able to extract orbital energies from a ``multiconfigurational'' wave function, in order to 
demonstrate the branching and energy changes of the orbitals, that is essential for the {\em primoid}\ second 
order Jahn-Teller cases. 

The optimization resulted in the following structure and energy:

\begin{Verbatim}
C4H4, D4h, GVB(PP)/def-TZVPP, units of Å and hartree

E=     -153.6925014955  GRAD. MAX=  0.0000280  R.M.S.=  0.0000162

  C           6.0  -0.7131801506  -0.7131801826  -0.0000000000
  C           6.0   0.7131801506  -0.7131801826   0.0000000000
  C           6.0  -0.7131801506   0.7131801826   0.0000000000
  C           6.0   0.7131801506   0.7131801826   0.0000000000
  H           1.0  -1.4687380037  -1.4687382006  -0.0000000000
  H           1.0   1.4687380037  -1.4687382006   0.0000000000
  H           1.0  -1.4687380037   1.4687382006   0.0000000000
  H           1.0   1.4687380037   1.4687382006   0.0000000000
\end{Verbatim}

The $D_{2h}$ minimum was calculated by choosing $D_{2h}$ symmetry in the input and slightly distorting the 
coordinates to some off-$x,y$-diagonal value but using the staring orbitals from the $D_{4h}$ run.
The optimization resulted in the following structure and energy:

\begin{Verbatim}
C4H4, D2h, GVB(PP)/def-TZVPP, units of Å and hartree

E=     -153.7126953359  GRAD. MAX=  0.0000191  R.M.S.=  0.0000099

  C           6.0  -0.7761786150  -0.6619033512   0.0000000000
  C           6.0   0.7761786150  -0.6619033512  -0.0000000000
  C           6.0  -0.7761786150   0.6619033512  -0.0000000000
  C           6.0   0.7761786150   0.6619033512  -0.0000000000
  H           1.0  -1.5315443480  -1.4196296797   0.0000000000
  H           1.0   1.5315443480  -1.4196296797  -0.0000000000
  H           1.0  -1.5315443480   1.4196296797  -0.0000000000
  H           1.0   1.5315443480   1.4196296797  -0.0000000000
\end{Verbatim}

\begin{figure}[H]
\centering
\includegraphics[width=17cm]{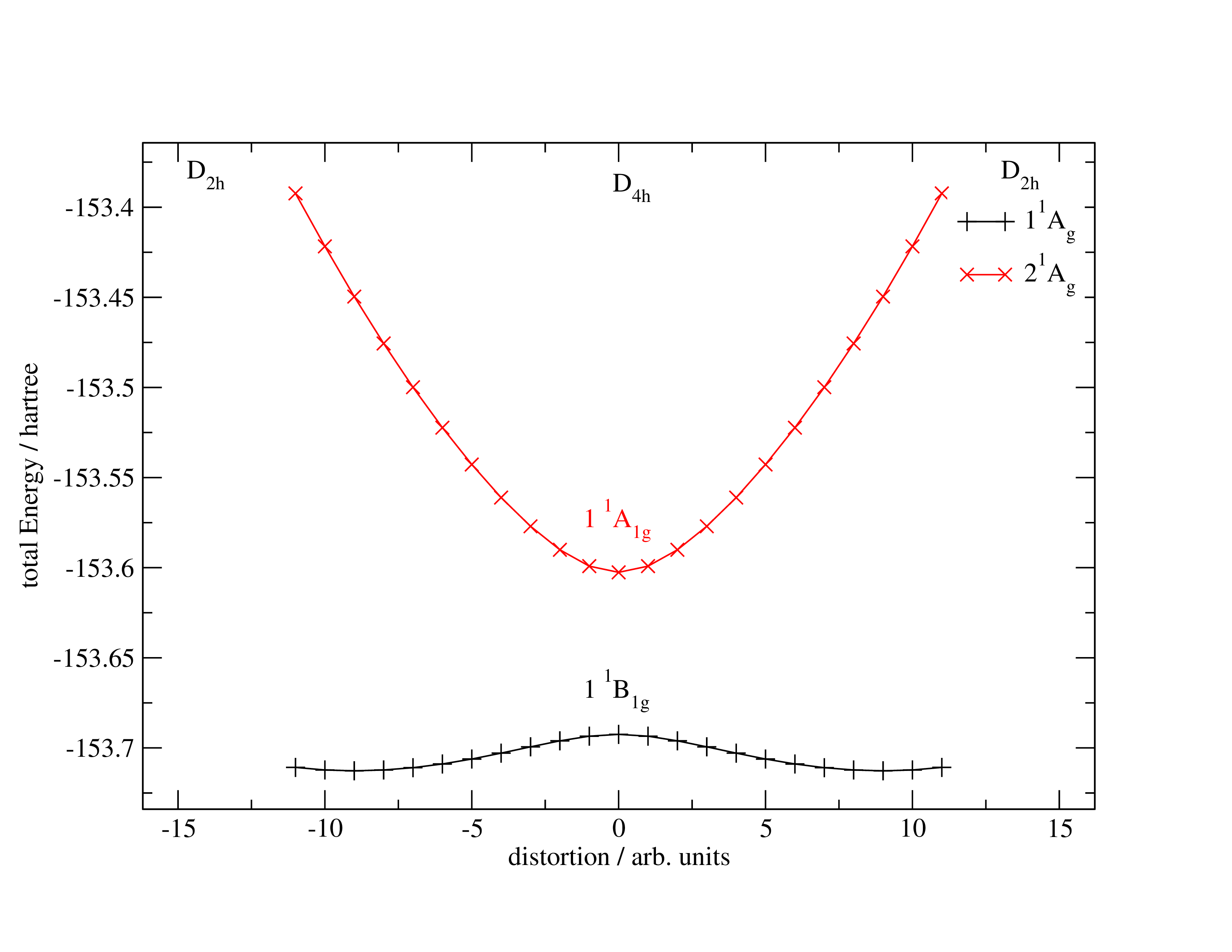}
\caption{Lowest two singlet states of C$_4$H$_4$ ($1^1B_{1g}$ and $1^1A_{1g}$) in $D_{4h}$ symmetry. 
Upon distortion to $D_{2h}$ they restrict to $1^1A_{g}$ and $2^1A_{g}$, respectively. 
Energies taken from GVB(PP)/def-TZVPP level of theory. The elevation to a (local) 
maximum at $x=0$ in the ground state level, can be interpreted as an effect of a second order Jahn-Teller 
coupling between the two levels by the distortional mode, which is represented by $B_{1g}$. \label{C4H4_orb_branch}}
\end{figure}

By a linear interpolation between the $D_{4h}$ saddle point and the $D_{2h}$ minimum structure 9 intermediate 
structures were generated and the frontier orbital energies were extracted a generate a branching plot 
(see Fig. \ref{C4H4_orb_branch}), a schematic version of which is represented in Fig. \ref{C4H4_levels} in 
the main text.

\begin{figure}[H]
\centering
\includegraphics[width=17cm]{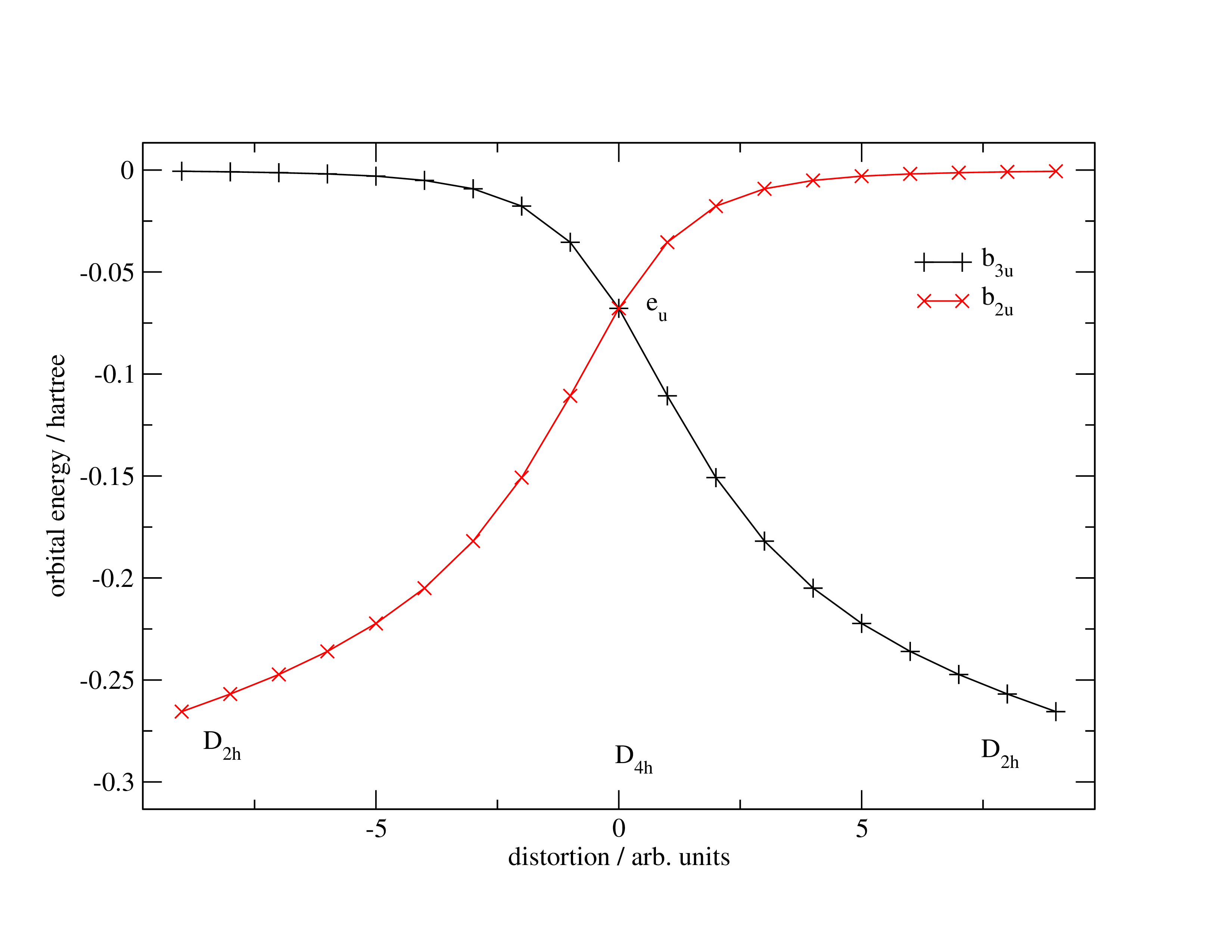}
\caption{Frontier orbital branching in C$_4$H$_4$ upon relaxed distortion from $D_{4h}$ to $D_{2h}$ at 
GVB(PP)/def-TZVPP level of theory.\label{C4H4_state_scan}}
\end{figure}

The corresponding state energies of first two singlet states ($1^1A_g$ and $2^1A_g$) have been computed for the 
intermediate structures complemented by two more strongly distorted structures by the analogous CAS(2i2) 
state average calculation. In this way one can see a representation of the JT-coupling between the two states 
(see Fig. \ref{C4H4_state_scan}). 

The orbital plots in Figures
\ref{C4H4-orbitals}, \ref{C8H8-orbitals} and \ref{MnF3_orbs} were generated with the program 
gabedit\cite{10.1002/jcc.21600}.

Magnetically induced currents were calculated for the $D_{2h}$ minimum structure 
using the program ReSpect (version 4.0.0) from Repisky and Komorovsky\cite{ReSpect}. 
Since for the $D_{2h}$ minimum structure the CI coefficients of the two configurations from the GVB(PP) 
calculation were  only 0.980137, and -0.198321 respectively (corresponding to a weight of only about 4\%
of the second configuration), a single reference method can be used safely for computation of the 
magnetic response. To stay consistent we have also used for the magnetic response the HF level of theory 
and the Dyalls (relativistic) triple-$\zeta$ basis sets for C and H\cite{10.1021/jp905057q}.

For the wave function calculation the GVB(PP) structure has been used a point nucleus model, the ``mdhf'' 
method (multicomponent Dirac-HF hamiltonian) but quasi-non relativistic settings ``cscale=20.0'' and 
``soscale=0.0'' and ``grid: large''.

The magnetic response calculation was done using the London orbital approach (GIAO) and the `
`dft-kernel: xalda'' option and the magnetic field was set perpendicular to the molecular plane.

The integration grid for the total current calculation was set to start -4 \AA\ below the molecular plane
pass the centre of the molecule and have a width and heights of 8 \AA , respectively and using 
$200\times 200$ grid points. 
The net currents passing that plane are calculated to 2.6748885 nA/T (diatropic) and 
-25.132882 nA/T (paratropic) yielding a total paratropic current of -22.457993 nA/T.

The streamline plot from Fig. \ref{C4H4_currents} were generated similarly just using a current plane starting 
at $x=-2.5$ and $y=-2.5$ \AA. The length and width was set to 5 \AA. The streamline plot was generated 
using the ``make-stream'' script that is distributed with ReSpect. Fig. \ref{C4H4_currents} was generated with 
Povray\cite{Povray}.

\subsection{C$_8$H$_8$}

Similar as C$_4$H$_4$ also C$_8$H$_8$ in full $D_{8h}$ symmetry is a multi-reference case of two 
electrons in the doubly degenerate $e_{2u}$ orbitals \cite{10.1021/jp8067365}, that has to be treated at least with 
a two-configuration wave function. For consistency we have applied here as well the GVB(PP)/TZVPP method with 
completely analogous strategy as in the case of C$_4$H$_4$, that is first optimization of the dianion using RHF 
and then followed by the symmetry restricted geometry optimization in D$_{8h}$ with GVB using 27 doubly occupied 
orbitals and two orbitals with two electrons in perfect pairing to yield a second order saddle point structure 
(with two equally weighted configurations):

\begin{Verbatim}
C8H8, D8h, GVB(PP)/def-TZVPP, units of Å and hartree

E=-307.5867677504

C       1.6802926394      0.0000000000      0.6960000000
C      -1.6802926394      0.0000000000      0.6960000000
C       1.6802926394      0.0000000000     -0.6960000000
C      -1.6802926394      0.0000000000     -0.6960000000
C      -0.6960000000      0.0000000000     -1.6802926394
C       0.6960000000      0.0000000000      1.6802926394
C      -0.6960000000      0.0000000000      1.6802926394
C       0.6960000000      0.0000000000     -1.6802926394
H       2.6743870164      0.0000000000      1.1077673732
H      -2.6743870164      0.0000000000      1.1077673732
H       2.6743870164      0.0000000000     -1.1077673732
H      -2.6743870164      0.0000000000     -1.1077673732
H       1.1077673732      0.0000000000      2.6743870164
H      -1.1077673732      0.0000000000      2.6743870164
H       1.1077673732      0.0000000000     -2.6743870164
H      -1.1077673732      0.0000000000     -2.6743870164






\end{Verbatim}

The C-C distance is 1.392 and H-H is 1.076 \AA , respectively, the energy of the HOMO 
$\varepsilon(e_{2u})=-0.0608$ Hartree.\\

Subsequent optimization in $D_{4h}$ results in a first order 
saddle point, with already a small coefficient of the second configuration 
($ c_1= 0.995999,\; c_2=-0.089369$) and HOMO and LUMO energies of $\varepsilon(b_{3u})=-0.2699$ and 
$\varepsilon(b_{2u})=0.0000$ Hartree, the C-C distances split into two sets of 1.324 and 1.473 \AA and C-H distances 
of 1.075 \AA: 

\begin{Verbatim}
C8H8, D4h, GVB(PP)/def-TZVPP, units of Å and hartree

E=-307.6196129777

C         1.6689419240     -0.0000000000      0.7365511863
C         1.6689419240     -0.0000000000     -0.7365511863
C        -1.6689419240     -0.0000000000      0.7365511863
C        -1.6689419240     -0.0000000000     -0.7365511863
C         0.7382750555      0.0000000000      1.6782476200
C         0.7382750555      0.0000000000     -1.6782476200
C        -0.7382750555      0.0000000000      1.6782476200
C        -0.7382750555      0.0000000000     -1.6782476200
H         2.6649503331      0.0000000000      1.1423860279
H         2.6649503331      0.0000000000     -1.1423860279
H        -2.6649503331      0.0000000000      1.1423860279
H        -2.6649503331      0.0000000000     -1.1423860279
H         1.1298611754      0.0000000000      2.6795220078
H         1.1298611754      0.0000000000     -2.6795220078
H        -1.1298611754      0.0000000000      2.6795220078
H        -1.1298611754      0.0000000000     -2.6795220078








\end{Verbatim}

Releasing the symmetry to $D_{2d}$ yields the minimum structure, with again smaller CI coefficient
for the second configuration ($c_1=0.999143,\; c_2=-0.041393$) and HOMO and LUMO energies of 
$\varepsilon(a_1)=-0.3578$ and $\varepsilon(a_2)= 0.0000$ Hartree; geometric parameters are C-C$_{short/long}= 1.320/1.475 $\AA , C-H$=1.077$ \AA , $\sphericalangle$(CCC)$=127.3^\circ$, $\sphericalangle^{dih}$(CCCC)$=0.0/54.1^\circ$. 

\begin{Verbatim}
C8H8, D2d, GVB(PP)/def-TZVPP, units of Å and hartree

E=-307.6408330022

C    -1.5659424    0.6328078    0.3790173 
C     1.5659424    0.6328078   -0.3790173 
C    -1.5659424   -0.6328078   -0.3790173 
C     1.5659424   -0.6328078    0.3790173 
C    -0.6328078    1.5659424    0.3790173 
C     0.6328078   -1.5659424    0.3790173 
C    -0.6328078   -1.5659424   -0.3790173 
C     0.6328078    1.5659424   -0.3790173 
H    -2.4628900    0.8232252    0.9441068 
H     2.4628900   -0.8232252    0.9441068 
H    -2.4628900   -0.8232252   -0.9441068 
H     2.4628900    0.8232252   -0.9441068 
H     0.8232252   -2.4628900    0.9441068 
H    -0.8232252    2.4628900    0.9441068 
H     0.8232252    2.4628900   -0.9441068 
H    -0.8232252   -2.4628900   -0.9441068 
\end{Verbatim}

The magnetically induced currents for the $D_{2d}$ minimum structure were calculated with
the very same procedures as in case of C$_4$H$_4$ and the magnetic field was set perpendicular to 
the average molecular plane. The integration grid for the total current 
calculation was set to start -5 \AA\ below the molecular plane
pass the centre of the molecule and have a width and heights of 10 \AA , respectively 
$300\times 300$ grid points. The net currents passing that plane are calculated to 5.7644395 nA/T (diatropic) and 
-8.2361711 nA/T (paratropic) yielding a total paratropic current of -2.4717316 nA/T. The streamline plot 
from Fig. \ref{C4H4_currents} were generated similarly just using a current plane starting 
at $x=6$ and $y=-6$ \AA. The length and width was set to 12 \AA.

\subsection{MnF$_3$}

For the calculation of the magnetically induced molecular currents in MnF$_3$, 
we have used the experimental gas-phase structure, that is $C_{2v}$ symmetric and corresponds
to the minimum of the $^5B_1$ electronic state\cite{10.1021/ja9712128} (Mn-F1$ = 1.728$\AA ,Mn-F2$ = 1.754$\AA , Mn-F3$ = 1.754$\AA  
and $\sphericalangle$(F2-Mn-F3)$ = 143.3^\circ$. We have used the PBE0 hybrid 
density functional\cite{doi:10.1063/1.472933} as it is implemented in ReSpect (``method : mdks/pbe0''), the 
non-relativistic settings as described above, ``multiplicity : 5'' and again the relativistic basis sets from Dyall for Mn 
and F atoms\cite{10.1021/jp905057q}. In the magnetic response calculation the DFT kernel option ``xalda'' was chosen.
The grid plane for the current vectors was set to the $z=0$ plane extending 8 \AA in $x$ and $y$ direction respectively with a 
grid of $50\times 50$ points. Otherwise the same procedure was applied as in case of C$_4$H$_4$ and C$_8$H$_8$.

\section{On the physical and chemical interpretation and connection of induced paramagneticity and bonds}

Yet a more generalizing way of view is that diamagnetism indicates situations
of electron delocalized bonds while paramagnetism indicates situations of incompletely filled
molecular shells. And adopting a purely physical point of view, one can formulate, that 
proper bonds have a tendency to undergo precession motions in the same way electrons would 
that are orbiting in quasi-classically stable trajectories, 
while for symmetry reasons instable trajectories from a near-degenerate trajectory space 
lead to precession motions within this space but in opposite direction as compared to 
the diamagnetic case.\\


The symmetry connection between rotational virtual excitations and the Jahn-Teller effect in a subfamily of the 
point groups deepens our understanding of antiaromaticity. On the one hand, this connection, 
however cannot be established in the general case of an antiaromatic molecule by pure symmetry 
considerations, not even in the case of a relatively high symmetry like pentalene. This is the usual 
restriction underlying any symmetry selection rule (take dipole transitions for electronic excitations by 
light). An open question in this context also is if there is a deeper connection between the enumerator and the 
denominator in expression \ref{vt}, which intuition might suggest: two orbitals that are very similar in shape and 
only differ by orientation necessarily should have similar energies. An open question is: In how far can we expect 
the opposite as well?

On the other hand the bare symmetry rule or numerical principle ({\em i.e.} the conditions under 
which to observe large terms \ref{vt}) again cannot provide us with a precise condition as to 
when a specific case is to be considered antiaromatic and when not. We even notice a paradoxical component: 
the stronger the distortion the smaller the terms in expression \ref{vt} are getting (by both an increase 
in the denominator and a decrease in the numerator) and the weaker the paramagnetic response will be in general. 
Also a strong paramagnetic response shall make one expect a low degree of distortion as well. In 
that sense we rather expect that all systems that meet the IUPAC definition (that without a doubt by construction 
is based on observed real world examples) are those that find a kind of balance between the extremes of 
distortion and paramagneticity. C$_8$H$_8$ misses it due to the strong distortion while there might be 
other cases of strong paramagneticity that show less or no distortion contrary to expectation. 
Take for example a transition state of a $2+2$ cycloaddition reaction.\\

Thus the concept of antiaromaticity appears as a principle of instability. Nevertheless its a concept that can be sensitively
verified (or falsified) both in the real world and in the computer by energetic, structural and magnetic methods.\\

\section{IUPAC definition of Antiaromaticity}

The definition of antiaromaticity which made it into the IUPACs ``Gold-Book'' was originally 
provided by V. I. Mishkin in his article about ``Glossary of terms used in theoretical organic chemistry''
\cite{10.1351/pac199971101919}. The complete quote is 
``antiaromaticity (antithetical to aromaticity)\\
Those cyclic molecules for which cyclic electron delocalization provides for the reduction (in some cases, loss) 
of thermodynamic stability compared to acyclic structural analogues are classified as antiaromatic species. In 
contrast to aromatic compounds, antiaromatic ones are prone to reactions causing changes in their structural 
type, and display tendency to alternation of bond lengths and fluxional behavior (see fluxional molecules) 
both in solution and in the solid. Antiaromatic molecules possess negative (or very low positive) values of 
resonance energy and a small energy gap between their highest occupied and lowest unoccupied molecular orbitals. 
In antiaromatic molecules, an external magnetic field induces a paramagnetic electron current. Whereas benzene 
represents the prototypical aromatic compound, cyclobuta-1,3-diene exemplifies the compound with most clearly 
defined antiaromatic properties.''

\bibliography{scibib}

\end{document}